%% file: pull-stream.tex
\RequirePackage{snapshot}
\documentclass[a4paper,UKenglish]{lipics-v2016}
 
\usepackage{microtype}
\usepackage{subcaption}

\title{A Formalization for Specifying and Implementing Correct Pull-Stream Modules}

\author[1]{Erick Lavoie}
\author[2]{Laurie Hendren}
\affil[1]{McGill University, 845 Sherbrooke Street West, Montreal, Quebec, Canada\\
  \texttt{erick.lavoie@mail.mcgill.ca}}
\affil[2]{McGill University, 845 Sherbrooke Street West, Montreal, Quebec, Canada\\
  \texttt{hendren@cs.mcgill.ca}}
\authorrunning{E. Lavoie and L. Hendren} 

\Copyright{Erick Lavoie and Laurie Hendren}


\EventEditors{}
\EventNoEds{0}
\EventLongTitle{Arxiv.org}
\EventShortTitle{Arxiv.org}
\EventAcronym{}
\EventYear{}
\EventDate{}
\EventLocation{}
\EventLogo{}
\SeriesVolume{}
\ArticleNo{} 

\begin{document}

\maketitle

\input{abstract}

\section{Introduction} 
\input{introduction}

\section{Background} 
\label{Section:Background}
\input{background}

\section{Insights and Approach}
\label{Section:Insights}
\input{insights}

\section{Event-Based Protocol Language}
\label{Section:Language}
\input{language}

\section{Pull-Stream Protocol}
\label{Section:Pull-Stream-Protocol}
\input{protocol}

\section{Reference Modules}
\label{Section:Model-Modules}
\input{model-modules}

\section{Evaluation of Community Modules}
\label{Section:Evaluation}
\input{evaluation}

\section{Related Work} 
\label{Section:RelatedWork}
\input{related}

\section{Conclusion and Future Work} 
\label{Section:Conclusion}
\input{conclusion}

\subparagraph*{Acknowledgements.}

We would like to thank Francisco Ferreira for his tremendous help in clarifying the presentation of our event-based protocol language and providing comments on drafts of this paper. This paper uses a modified version of the LIPICS v2016 template (\url{http://www.dagstuhl.de/en/publications/lipics/}).

\newpage
\appendix

\input{appendix}


\newpage
\bibliography{bibfile.bib}

\end{document}

%% file: abstract.tex
\begin{abstract}
\textit{Pull-stream} is a JavaScript demand-driven functional design pattern based on callback functions that enables the creation and easy composition of independent modules that are used to create streaming applications. It is used in popular open source projects and the community around it has created over a hundred compatible modules. While the description of the pull-stream design pattern may seem simple, it does exhibit complicated termination cases. Despite the popularity and large uptake of the pull-stream design pattern, there was no existing formal specification that could help programmers reason about the correctness of their implementations. 

Thus, the main contribution of this paper is to provide a formalization for specifying and implementing correct pull-stream modules based on the following: (1) we show the pull-stream design pattern is a form of declarative concurrent programming; (2) we present an event-based protocol language that supports our formalization, independently of JavaScript; (3) we provide the first precise and explicit definition of the expected sequences of events that happen at the interface of two modules, which we call the pull-stream protocol; (4) we specify reference modules that exhibit the full range of behaviors of the pull-stream protocol; (5) we validate our definitions against the community expectations by testing the existing core pull-stream modules against them and identify unspecified behaviors in existing modules.

Our approach helps to better understand the pull-stream protocol, to ensure interoperability of community modules, and to concisely and precisely specify new pull-stream abstractions in papers and documentation.
 \end{abstract}

%% file: introduction.tex
\textit{Pull-stream}~\cite{pull-stream} is a JavaScript demand-driven functional design pattern based on callback functions that enables the creation and easy composition of independent modules that are used to create streaming applications, initially proposed by Dominic Tarr~\cite{pull-stream-original-blog-post}. It is simple, because it does not require language support other than higher-order functions, yet it is rich enough to provide flow-control and correct termination behavior in case of errors. It also simplifies factoring complex applications into simpler reusable modules. It has already shown its worth by being used in the implementation of a data dissemination protocol for new social applications (\texttt{ssb}~\cite{secure-scuttlebutt}), in the JavaScript implementation of a peer-to-peer networking stack (\texttt{js-libp2p}~\cite{js-libp2p}), in a JavaScript implementation of a peer-to-peer hypermedia protocol (\texttt{js-ipfs}~\cite{js-ipfs}), and by being widely downloaded on the \texttt{npmjs} website\footnote{At the time of writing, the library had been downloaded over 90,000 times in the previous month. We believe most of these downloads are from small personal or custom in-house tools.}. Furthermore, an open-source community has grown around it and produced more than a hundred compatible pull-stream modules~\cite{pull-stream-community-modules}.

While the description of the pull-stream design pattern may seem simple, it does exhibit complicated termination cases. For example, we examined and tested the core pull-stream library~\cite{pull-stream} that has been under development for 5 years now and found two cases of unspecified behaviors~\cite{pull-stream-unspecified-behaviors}. While both are not major issues, and seem to not have created interoperability problems so far, their existence in a well-used library does show that even a seemingly simple callback protocol can have unexpected corner cases.
 
Despite the popularity and large uptake of the pull-stream design pattern, there was no existing formal specification that could help programmers reason about the correctness of their implementations. Thus, the main contribution of this paper is to provide a formalization for specifying and implementing correct pull-stream modules. 

We arrived at our current results in multiple steps. We first experimented by reimplementing some pull-stream modules in Oz~\cite{henz1993oz,van2004concepts}, a language with native stream support, to see if modules would be easier to implement and reason about in it. While it gave us insights about the nature of the pull-stream design pattern, the language is not well known and it was hard to explain the insights to a more general audience.  We therefore decided to provide a notation with a small number of rules that could capture those insights yet would be independent of both JavaScript and Oz. We then asked questions to the pull-stream community about the expected sequences of events that happen at the interface of two pull-stream modules, which we call the \textit{pull-stream protocol}. This way we identified all valid sequences of events and concisely captured the constraints in our notation. We then used the same notation to specify reference modules that use the full capabilities of the pull-stream protocol. They give a concise, precise, and complete reference of expected behavior, and can be used to test other module implementations. We finally validated our understanding on community contributed modules by implementing our reference modules in JavaScript. Using these modules allowed us to automatically discover unspecified behaviors in well-used modules. This formalization effort therefore helped clarify the expected behavior of pull-stream modules, should help module maintainers to ensure all community modules are inter-operable in the future, and provides a notation for concisely presenting new pull-stream abstractions in modules' documentation and future papers.

\subsection{Contributions}
In this paper we therefore make the following contributions:
\begin{itemize}
    \item we show how the pull-stream protocol implements an implicit stream of single-assignment dataflow variables using callbacks and how the benefits of a declarative concurrent programming model also apply to the pull-stream design pattern (Section~\ref{Section:Insights});
    \item we present an event-based protocol language that is used both to describe the pull-stream protocol and specify the behavior of pull-stream modules, independently of JavaScript (Section~\ref{Section:Language});
    \item we provide the first precise and explicit definition of the expected sequences of events that follow the pull-stream protocol at the interface of modules (Section~\ref{Section:Pull-Stream-Protocol});
    \item we specify parameterized modules that exhibit the full range of behaviors of the pull-stream protocol and can be used as references for implementations and for testing other modules (Section~\ref{Section:Model-Modules});
    \item we evaluate the conformity of community-contributed modules against our definitions to ensure the definitions adequately describe community expectations and can be used to find modules that do not correctly implement the protocol in all cases (Section~\ref{Section:Evaluation}).
\end{itemize}

To provide the necessary context, we first introduce the pull-stream design pattern using its JavaScript implementation (Section~\ref{Section:Background}). We then present the previous contributions in the aforementioned sequence. We then compare our work to the existing literature (Section~\ref{Section:RelatedWork}). We finally conclude with a brief recapitulation of our contributions and some future research directions (Section~\ref{Section:Conclusion}).

%% file: background.tex
\begin{figure}[htbp]
    \begin{center}
    \includegraphics[width=0.8\textwidth]{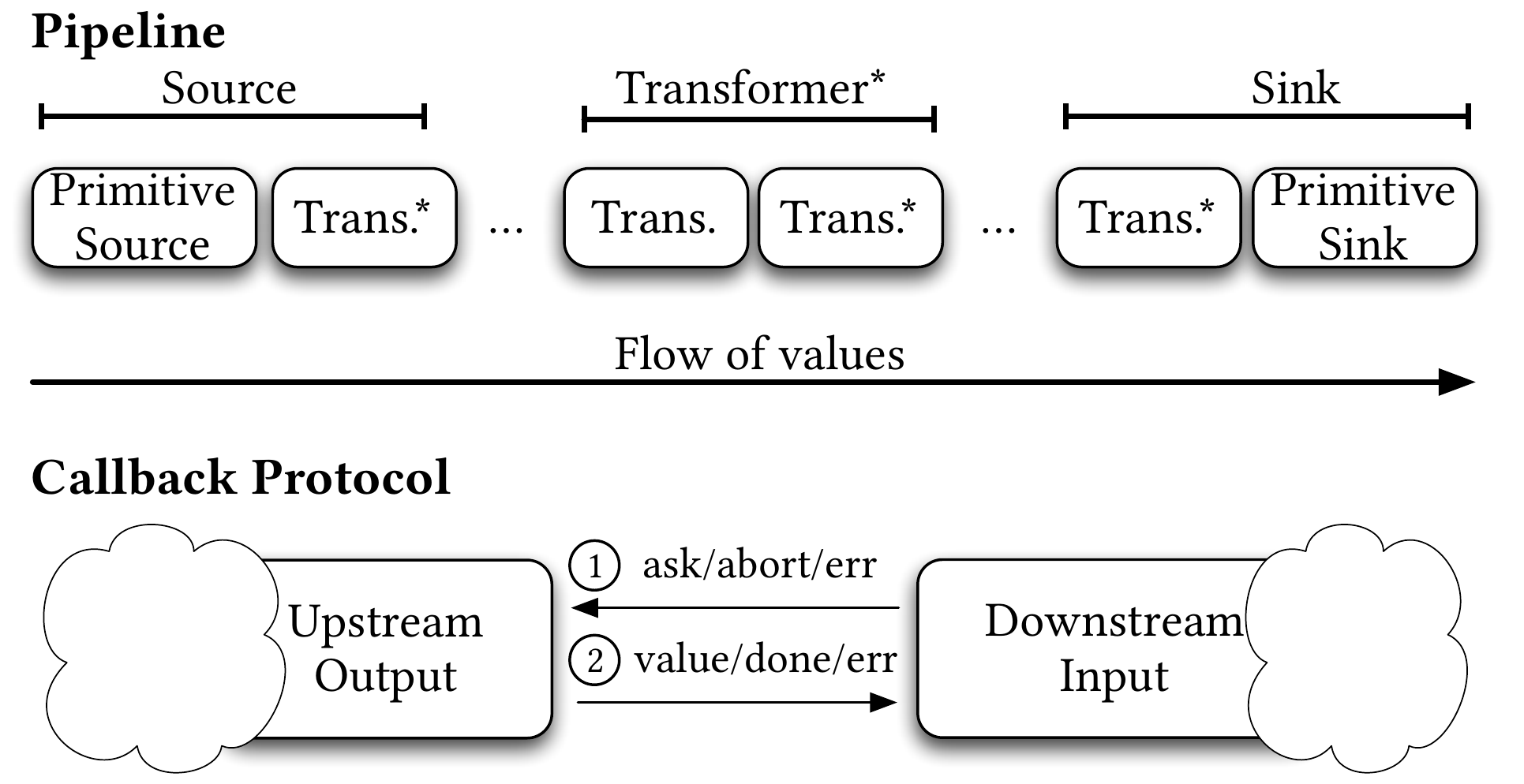}
    \end{center}
    \caption[Pull-stream design pattern.]{\label{fig:pull-stream-design-pattern} Pull-stream design pattern: pipeline of composable modules on top and callback protocol at the bottom.}
\end{figure}

The pull-stream design pattern is illustrated in Figure~\ref{fig:pull-stream-design-pattern}. It consists of both a composition mechanism to assemble individual modules in a pipeline and a callback protocol for enabling adjacent modules to communicate.

A pipeline is composed of three types of modules: a single source that produces values, a series of zero or more transformers\footnote{The existing documentation uses the name \textit{through} for transformers. The original designer later mentioned that he would have preferred transformer but stuck with the original name because the community adopted it. We break community conventions here to favor clarity.} that modify those values, and a single sink that consumes the values. The composition of multiple transformers is itself a valid transformer. Likewise, the composition of a transformer with a source or a sink is itself a valid source or sink. The stream values flow from left to right, from the source to the sink.

Adjacent modules communicate with a two-parameters callback protocol. The downstream module first makes a request by invoking the output function of the upstream module with a callback. Then, the upstream module replies with an answer by invoking the callback. The first parameter of either function determines the type of operation: the type of a request is determined by the \texttt{abort} parameter and the type of an answer is determined by the \texttt{done} parameter. There are therefore multiple cases to consider. 

In the normal and common case, a request \textit{asks} for a value by invoking the output function with \texttt{abort} set to \texttt{false} and a callback for the expected answer as a second parameter. An answer then returns a value by invoking the callback with the \texttt{done} parameter set to \texttt{false} and the value provided as a second parameter. 

In addition to the normal case, a request may \textit{abort} processing normally by invoking the output function with \texttt{abort} set to \texttt{true}, or abort abnormally with an error with \texttt{abort} set to \texttt{new Error(...)} (which is also truthy\footnote{In JavaScript, thruthy values can be used as a true value in conditional statements or expressions.}).

An answer may also \textit{terminate} the stream normally by setting the first parameter \texttt{done} to \texttt{true}, or abnormally with an error by setting the first parameter \texttt{done} to \texttt{new Error(...)}. 

\subsection{Example Module Implementations}
\begin{figure}[htbp]
    \centering
    \begin{subfigure}[t]{0.40\textwidth}
        \includegraphics[width=\textwidth]{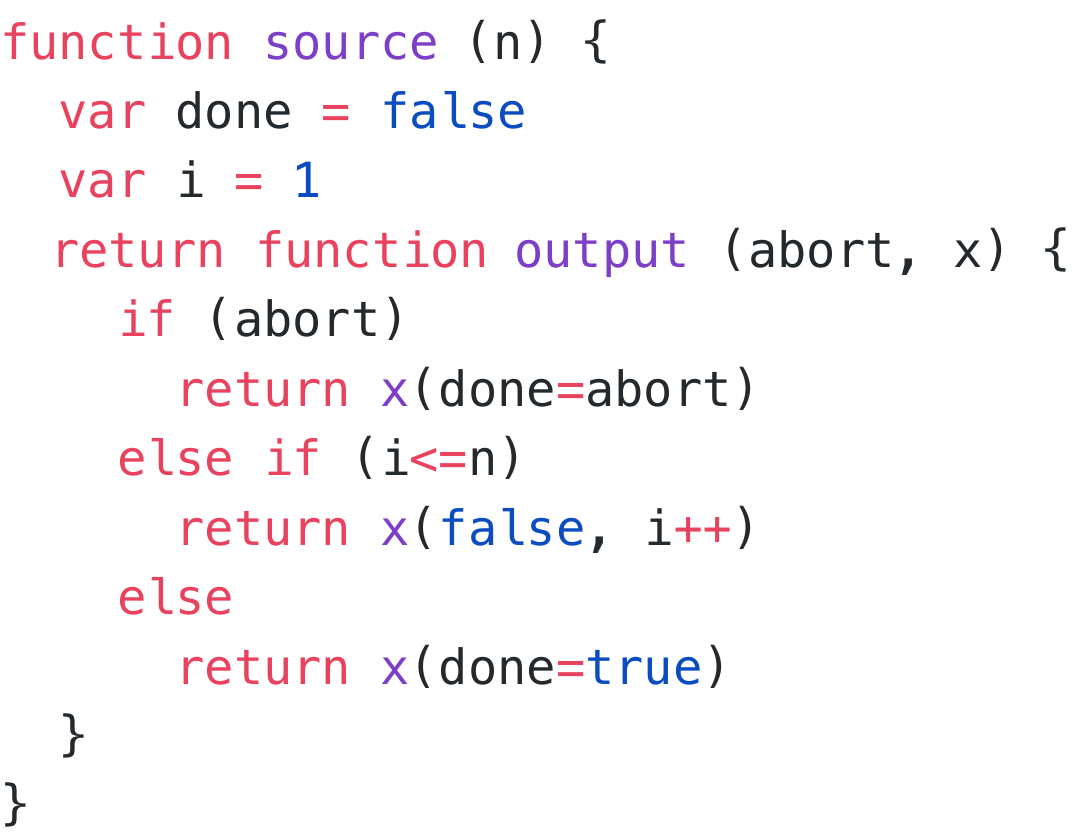}
        \caption{\label{fig:sources} Source example.}
    \end{subfigure}
    ~ 
    \begin{subfigure}[t]{0.52\textwidth}
        \includegraphics[width=\textwidth]{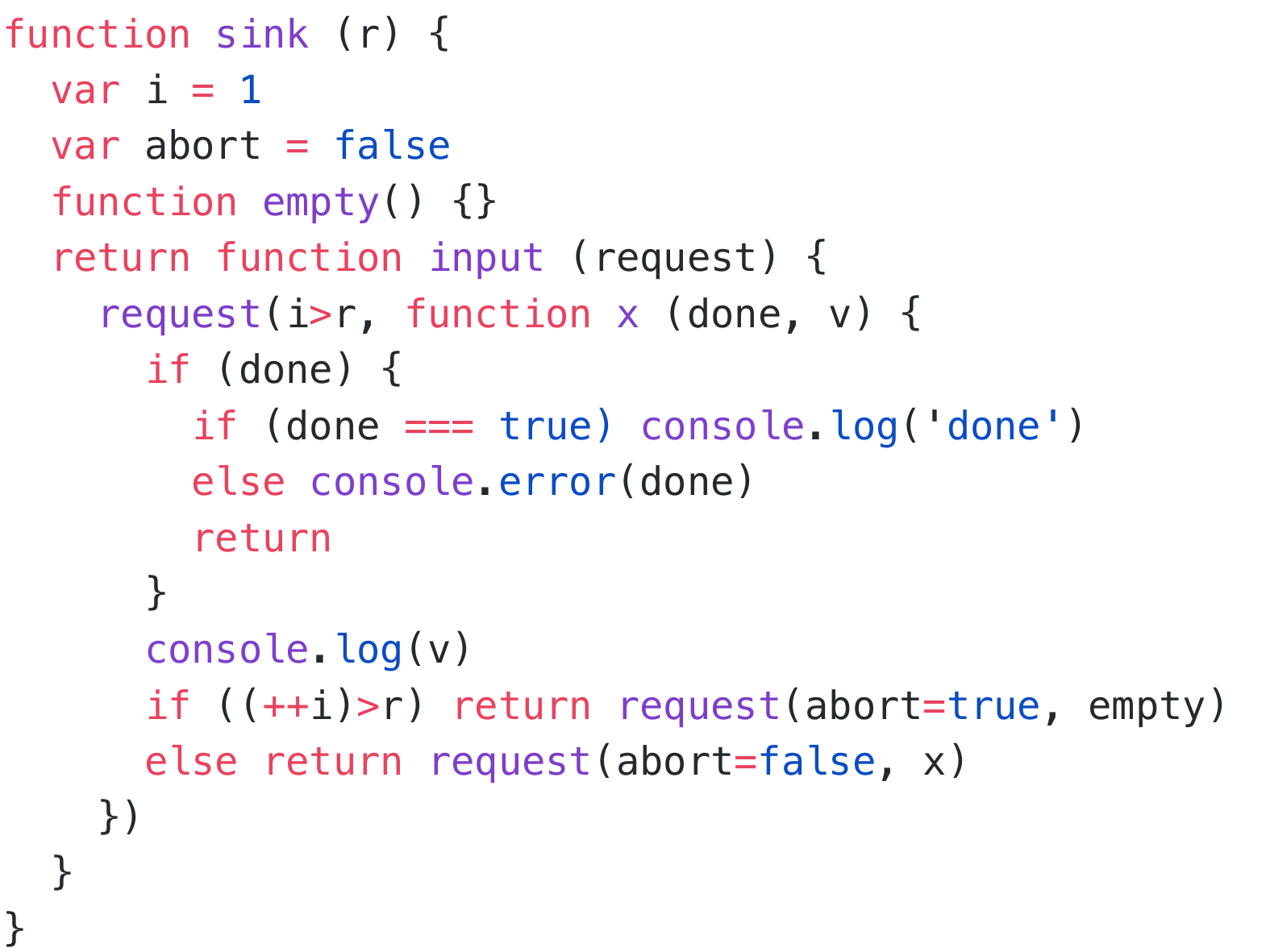}
        \caption{\label{fig:sinks} Sink example with aborting support after $r$ requests.}
    \end{subfigure}
    \caption[Source and Sink Examples.]{Source and Sink: the $request$ parameter of the sink function is the $output$ function of the module upstream (source or transformer). The abort flag is made explicit so the inverted logic of the protocol is easier to read.}
\end{figure}

The following modules illustrate the key features of the pull-stream protocol. An example source of values, that counts from $1$ to $n$, is implemented in Figure~\ref{fig:sources}. Instantiating the module returns a function named \textit{output}. A request is performed by invoking the output function with an abort flag and a callback function $x$. If the source is aborted from downstream (\texttt{abort} is \texttt{true} or an error), done will be set to the abort value and $x$, if defined, is called with the same value. This case is used by the module downstream to abort $early$, before all values have been output. Otherwise, if there are still values to output, $x$ is called with the current value (\texttt{done=false}). This is the normal case where a value flows from the output of an upstream module to the input of a downstream module. Finally, in the last case, no more values are available and $x$ is called with $done$ (\texttt{done=true}). This is the normal termination case, where a source is allowed to output all its values and complete. This source example does not raise errors and therefore the third answer case is not illustrated.

An example of a sink, the complement of a source, is implemented in Figure~\ref{fig:sinks}. The sink takes an $r$ parameter to define the number of non-abort requests to perform. Instantiating the module returns an $input$ function. Invoking the input function with an output function as argument $connects$ both. The sink then $requests$ values from the upstream module by calling its $output$ function\footnote{Functions are both objects (nouns) and represent actions (verbs) that are initiated from outside the module. The existing documentation on pull-streams sometimes name the output function \textit{source} and sometimes \textit{read}. When we refer to the object that returns values from inside a module, we call it an \textit{output} function. When we refer to the action of obtaining values from that object from outside the module, we write \textit{requesting} a value and name the function a \textit{request} function.}, hence the parameter is named $request$. Once the input function is invoked, it starts making requests immediately. If $0$ requests are demanded, then the output function is \textit{aborted}. Otherwise, a new value is $asked$ (\texttt{abort=false}). In both cases, the callback $x$ is passed to obtain an answer. The module then waits for an answer to be provided and therefore for $x$ to be invoked. If the source has completed or has failed, \texttt{'done'} or an error is printed on the console. If a new value is returned then it is printed on the console and a new request is made if some are left. Since requests are initiated from the sink, the protocol is demand-driven and lazy: a new value is not produced until one has been explicitly $requested$.

\begin{figure}[htbp]
    \centering
    \begin{subfigure}[t]{0.48\textwidth}
        \includegraphics[width=\textwidth]{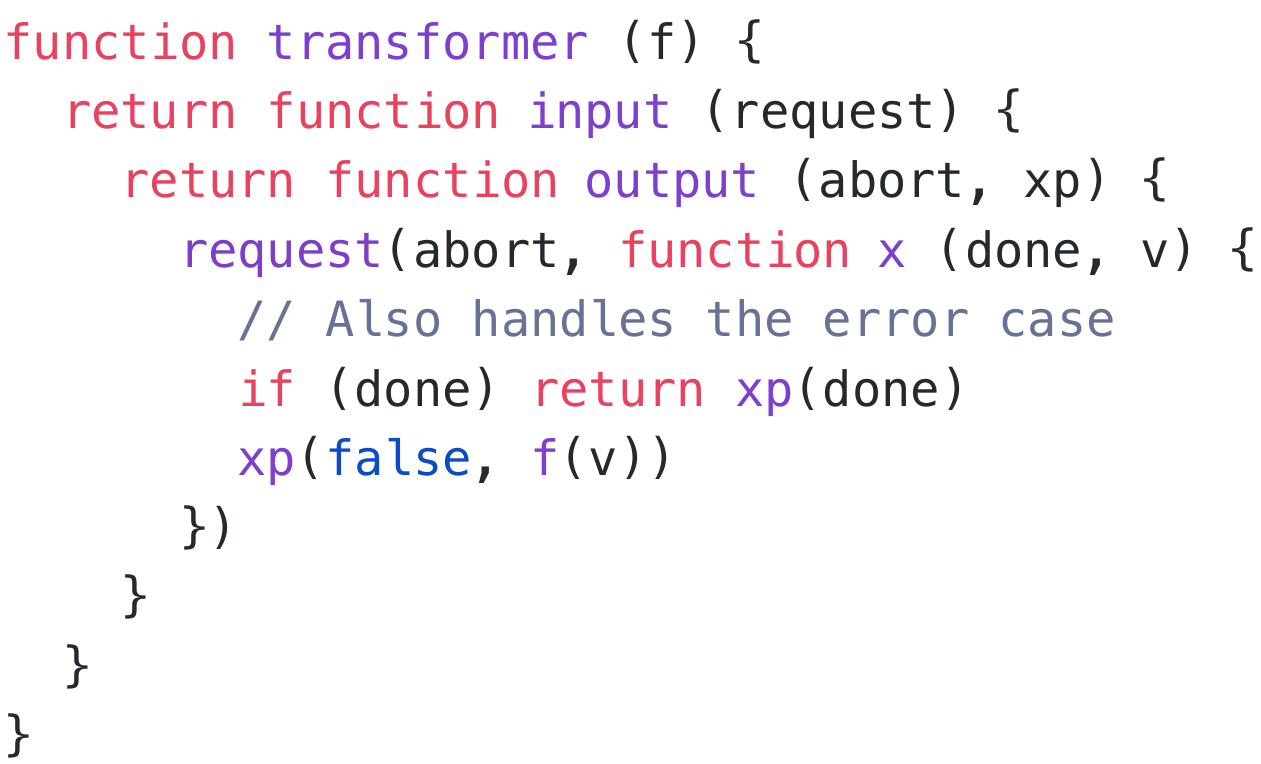}
        \caption{\label{fig:transformers} Transformer example.}
    \end{subfigure}
    ~ 
    \begin{subfigure}[t]{0.48\textwidth}
        \includegraphics[width=\textwidth]{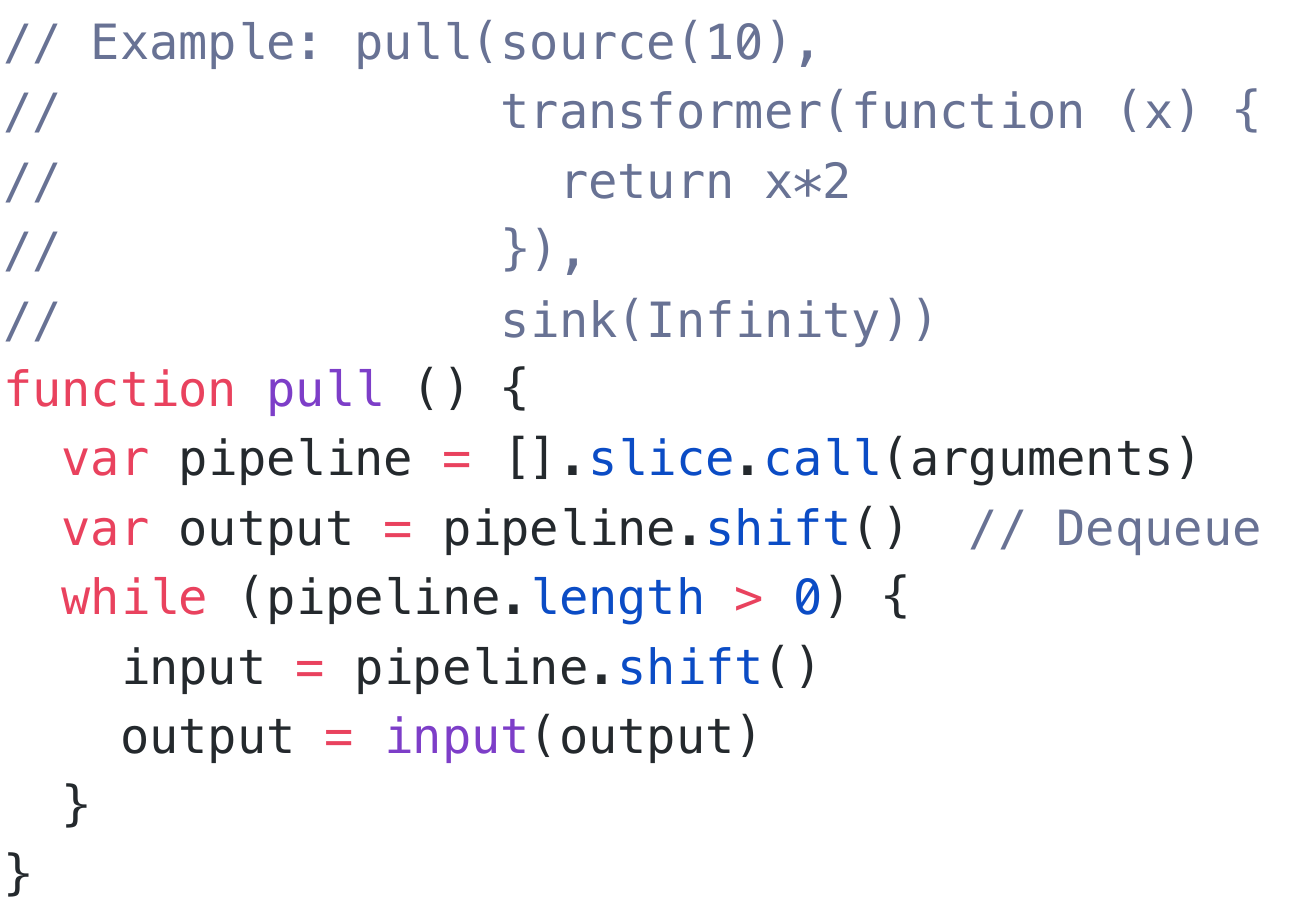}
         \caption{\label{fig:connect}  Pull helper function to create the stream pipeline.}
    \end{subfigure}
    \caption[Transformer and Pull Examples.]{Transformer and Pull: in JavaScript, \texttt{arguments} contains all the call-site arguments regardless of the function definition. Moreover, it behaves like an array but does not have all its methods therefore it is converted to an array using the reflection API (\texttt{call} on the slice method of an array that produces the \texttt{pipeline} value).}
\end{figure}

An example of a transformer, which takes input values from upstream, applies a function $f$ on them, and outputs the results downstream, is implemented in Figure~\ref{fig:transformers}. It combines both an $input$ and an \textit{output} function. The module is instantiated with a single-parameter function $f$ which outputs a result when given an input value. It returns an \textit{input} function, that expects an \textit{output} function as a parameter, similar to a sink. Once invoked, the input function returns a new \textit{output} which may be used as a source. Passing a source to the input function of a transformer therefore returns a new source. The output function of the transformer, directly forwards its \textit{requests} to the upstream module, including the abort cases but with a different callback $x$ rather than $xp$, to process the incoming value. It then waits for an answer until $x$ is invoked. If the upstream module is done or has failed, it forwards the answer downstream. Otherwise, it applies $f$ on the value $v$ and pass the result downstream by invoking $xp$.

Other modules, such as bi-directional network sockets may also have both an input and an output. On one side of the communication channel they can be used as a transformer and on the other as both a source and a sink. In either case, their behavior is similar to the three previous cases shown.

An entire pipeline may be connected by passing the output function of a module upstream to the input function of the next module downstream. The process is illustrated in Figure~\ref{fig:connect}. The actual implementation\footnote{\url{https://github.com/pull-stream/pull-stream/blob/master/pull.js}} is a bit more complicated. It allows any possible combination of modules that is not a full pipeline to return a source, a transformer, or a sink module than can be reused later. It also allows modules to be defined in object form, in which the input and output functions are methods\footnote{Respectively named source and sink. We prefer input and output because the first letter of each is different and makes it easier to identify the \textit{ports} later in the formalism.}.

\subsection{Pull-Stream Design Pattern Properties}

The pull-stream design pattern provides a combination of many interesting properties:
\begin{itemize}
	\item An upstream module (producer) and a downstream module (consumer) may both regulate the flow of values by respectively delaying the current answer and the next request;
	\item The consumer may abort the stream early even though the producer may still have more values to provide;
    \item Errors are handled within the protocol;
    \item Any module may propagate an error and has an opportunity for cleaning up after an error or the termination of the stream;
    \item The values are generated lazily therefore a source may produce infinitely many values;
    \item Modules may be composed before the construction of the complete pipeline which favors reuse of code when building libraries;
    \item Both the composition of modules and the construction of a pipeline is declarative: it does not require an understanding of the callback protocol by the users of modules;
    \item The implementation of modules may use concurrency to improve the overall throughput (ex: it may request multiple values and process them in parallel before returning its results). Outputs may or may not be in order.
\end{itemize}

%% file: insights.tex

To better understand the pull-stream design pattern, we implemented some pull-stream modules in the Oz language. We explain here the insights we obtained from the experience which in turn informed our formalization approach.

\begin{figure}[htbp]
    \centering
    \begin{subfigure}[b]{0.55\textwidth}
        \includegraphics[width=\textwidth]{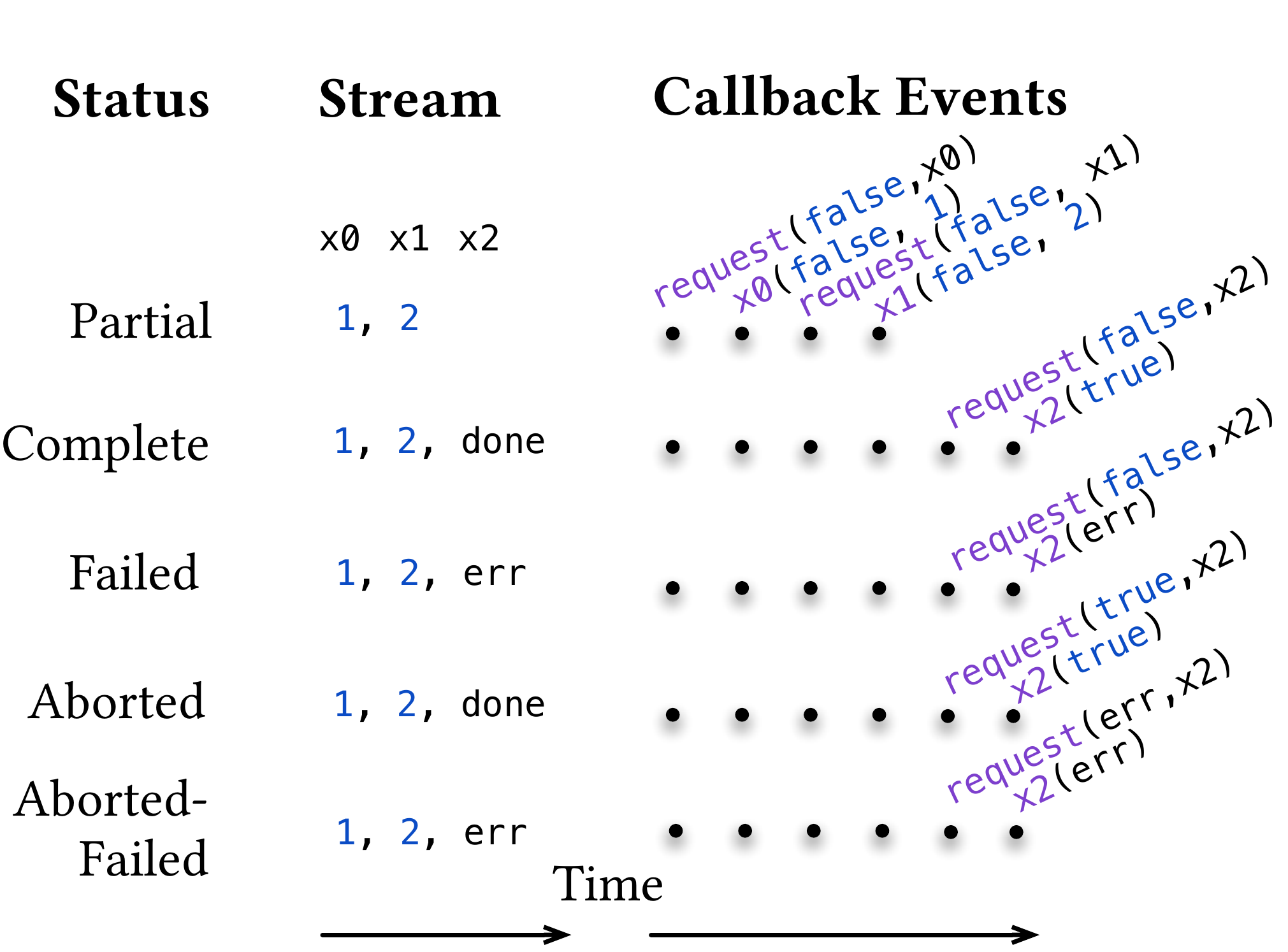}
        \caption{\label{fig:callback-stream} Implicit callback streams.}
    \end{subfigure}
    ~ 
    \begin{subfigure}[b]{0.4\textwidth}
\begin{tabular}{ll}
\textbf{Syntax}                    & \textbf{JavaScript Events}              \\
$ask[\bar x_i]$                    & \texttt{request(false, xi)}             \\
$abort[\bar x_i]$                  & \texttt{request(true, xi)}              \\
$error[err, \bar x_i]$             & \texttt{request(err, xi)}               \\
$x_i := v_i$                       & \texttt{xi(false, vi)}                  \\
$x_i := done$                      & \texttt{xi(true)}                       \\
$x_i := err$                       & \texttt{xi(err)}                        \\
\end{tabular}
\caption{Formal equivalent.}
\label{tb:js-events}
    \end{subfigure}
    \caption[Callback events.]{Callback events that form implicit streams and introduction to their equivalent formal representation.}
\end{figure}

Our key insight is that the sequence of callbacks at the interface of two modules creates an implicit stream. We may view this stream as a stream of \textit{single-assignment dataflow variables}, as illustrated in Figure~\ref{fig:callback-stream}. Invoking the output function \textit{extends} the stream with a new variable and invoking a callback $binds$ the value of that variable. As each callback is invoked only once, the variables are assigned only once. Since the behavior of modules is triggered by callback events, the assignment of variables can be used for synchronization as in dataflow programming. 

Programming with concurrent streams of single-assignment dataflow variables is a form of \textit{declarative concurrency}\footnote{Chapter 4 of \textit{Concepts, Models, and Techniques of Computer Programming} by Van Roy and Haridi~\cite{van2004concepts} provides an exhaustive discussion of declarative concurrency.}, with Unix pipes probably being the best-known example of the programming model. This model makes reasoning about concurrent applications easier than other concurrent models because the non-determinism in the concurrent events is not observable: regardless of the concrete execution order, the result is always the same. In other words, a declarative concurrent streaming program produces the same stream output, regardless of the exact order in which the individual values have been computed. Abstractions built within that model bring the benefits of parallel processing without the complexity of reasoning about multiple orders of executions. Moreover, similar to the composition of functions which is itself a function, the composition of declarative concurrent streaming modules is itself declarative concurrent. It allows complex programs to be easily built from simpler modules.

Some of the most complex pull-stream modules involve managing multiple streams concurrently, \texttt{pull-many}~\cite{pull-many} and \texttt{pull-lend-stream}~\cite{pull-lend-stream} being two examples. Nonetheless, they are still easy to use because they are declarative concurrent. While their usage is simple, their implementation is not. Unless all possible execution cases are correctly handled, the implementation may break the declarative concurrency model in some cases. Moreover, because there are multiple termination cases in the pull-stream protocol, the correct termination of the implementation is non-trivial to establish for all of them. Both concurrency and termination need to be accounted for to provide correct implementations, therefore the notation we introduce later supports specifying concurrent events and behaviors.

The JavaScript execution model is single-threaded. However, the \textit{asynchronous} execution of some of the libraries available in the execution environment, such as the Document Object Model (DOM) functions in a browser or the input-output functions in Node.js, may happen in parallel. This means that the results, usually obtained in callbacks, may arrive in any order. This is precisely what the inter-leaving semantics\footnote{\textit{Idem}, Section 4.1.1} for concurrent programs models captures and therefore fits nicely with the actual programming model programmer use when programming JavaScript applications.

Figure~\ref{tb:js-events} provides a preview of the equivalence between the events that happen in the JavaScript implementation we presented in Section~\ref{Section:Background} and the syntax we use in the rest of the paper. The next section introduce it more formally.

%% file: language.tex
\newenvironment{rcases}
  {\left.\begin{aligned}}
  {\end{aligned}\right\rbrace}
  
\newenvironment{rsqcases}
  {\left.\begin{aligned}}
  {\end{aligned}\right\rbrack}
  
\makeatletter
\newenvironment{sqcases}{%
  \matrix@check\sqcases\env@sqcases
}{%
  \endarray\right.%
}
\def\env@sqcases{%
  \let\@ifnextchar\new@ifnextchar
  \left\lbrack
  \def\arraystretch{1.2}%
  \array{@{}l@{\quad}l@{}}%
}
\makeatother

In this section we present the notation we use in the rest of the paper. The main goal of our notation was to provide a precise and concise notation to specify the pull-stream protocol and new pull-stream modules in papers that would be independent of JavaScript. Surprisingly to us, the notation we obtained allowed us to avoid the description of the internal state of modules, which makes it usable with many languages, by a more general audience than the original JavaScript community that currently use it.

\subsection{Syntax}
\label{Section:Syntax}

\newcommand{\ebnfname}[1]{\ensuremath{\text{\textit{#1}}}}

The syntax of our notation is described using the Extended Backus-Naur Form~\cite{iso-14977}. We organize the rules in groups and explain them in sequence. In addition to the usual symbols, we sometimes use mathematical syntax elements such as overline (ex: $\overline{x}$), underline (ex: $\underline{x}$), and subscripts (ex: $x_2$). 

Table~\ref{tb:syntax-basic-elements} shows the syntax of basic elements of the notation. \textit{boolean}, \textit{number}, \textit{letter}, \textit{alphanumeric}, \textit{name}, and \textit{variable} have common syntaxes. In addition, we use a particular syntax for stream related concepts. The \textit{stream-index} represents the position of a variable or a value in a stream. It may either be a number that represents a concrete single position in a stream (ex: $1$ represents the first position), or a variable that can represent \textit{any} position in a stream. A \textit{stream variable} represents a \textit{single-assignment dataflow variable}, which in JavaScript is implemented with a callback function that should be called only once. A \textit{stream value} represents a value the variable takes once it is bound. The \textit{stream complete} symbol is a special variable value that signifies the stream has completed and has no more values. The \textit{stream failed} symbol is another special variable value that signifies that the stream has failed with an error. 

\begin{table}[htbp]
\centering
\begin{tabular}{ll}
\textbf{Syntax}                                & \textbf{Examples}                                \\
$boolean = "\top" \mid "\bot";$
                                               & $\top$ (true), $\bot$ (false)
\\
$number = digit, \{digit\};$                    & $0~1337$                                         \\
$letter = character;$                           & $a~b~z$                                          \\
$alphanumeric = digit \mid letter;$             & $0~a$                                            \\                          
$name = letter, \{alphanumeric\};$              & $ask~abort$                                      \\                                       
$\ebnfname{variable} = letter, \{letter\};$    & $n~r~te$                                         \\
$\begin{aligned}
\ebnfname{stream-index} & = \ebnfname{variable} \mid number; 
\end{aligned}$                                 & $i~1~23$                                         \\
$\ebnfname{stream-variable} = 
\overline{letter}_{\ebnfname{stream-index}};$  
                                               & $\bar x_i~\bar x_1$                    \\
$\ebnfname{stream-value} = 
"v"_{\ebnfname{stream-index}};$  
                                               & $v_i$                                       \\
$\ebnfname{stream-complete} = "d","o","n","e";$ & $done$                                           \\
$\ebnfname{stream-failed} = "e","r","r"_{[number]};$ 
                                               & $err~err_1$                                      \\
\end{tabular}
    \caption[Syntax of basic elements.]{Syntax of basic elements in EBNF (also using overline, underline, and subscript syntax).}
\label{tb:syntax-basic-elements}
\end{table}

We use some conventions to make the notation easier to read, as listed in Table~\ref{tb:syntax-conventions}. We use $i$ and $j$ to represent stream indexes, quotes on variables and values (ex: $\bar x_1', v_i'$) to represent respectively the variables and values in a stream that have gone through a single stage of transformation. Any other letters are used to represent function parameters.

\begin{table}[htbp]
\centering
\begin{tabular}{lll}
\textbf{Kind}         & \textbf{Convention}  & \textbf{Example} \\
Stream index var.     & $"i" \mid "j"$       & $\bar x_i, v_j$ \\
Transformation stage  & $variable^{\{"'"\}}$ & $\bar x_i \rightarrow \bar x_i', v_i \rightarrow v_i'$\\
Function parameter    & other letter(s)      & $n, r, te$ \\
\end{tabular}
    \caption[Conventions.]{Conventions used for variables and values.}
\label{tb:syntax-conventions}
\end{table}

The basic elements and syntactic conventions are used to represent events, themselves listed in Table~\ref{tb:syntax-events}. This is useful to present the semantics rules later. An \textit{empty-event} is a special symbol used to express ordering rules without knowing what the target event is going to be already, whereas  \textit{arguments} are used in the syntax of other events and may either be a \textit{stream value}, a \textit{stream variable}, or a \textit{stream failed}. A \textit{method call} is an event that represents calls to a method of a pull-stream module that is not part of the base protocol. For example, some parameters of modules may be dynamically changed during execution and the method call represents when that happened and with which arguments. \textit{request} and \textit{answer} events are the basic events of the pull-stream protocol. As explained in Section~\ref{Section:Background} and Table~\ref{tb:js-events}, a request corresponds to the \texttt{request} function call, and an answer corresponds to the invocation of the callback provided in the request. A \textit{port} represents where an event is initiated, such as a downstream module in the case of a request. Ports enable the description of complex modules which interact with multiple streams at a time. An \textit{event} is a port and one of the other event types mentioned. A \textit{history} is a sequence of events separated by commas and represent \textit{all} the concrete events that happened during an execution.

\begin{table}[htbp]
\centering
\begin{tabular}{ll}
\textbf{Syntax}                                & \textbf{Examples}                                \\
$\ebnfname{empty-event} = "e","m","p","t","y";$ 
                                               & $empty$                                          \\

$\ebnfname{argument} = \ebnfname{stream-value} \mid \ebnfname{stream-variable} \mid \ebnfname{stream-failed};$
                                               & $v_i~\bar x_i~err$                               \\
$\ebnfname{arguments} = argument, \{"," argument\};$ 
                                               & $v_i~err,\bar x_i$                               \\
$\ebnfname{method-call-event} = name_{\ebnfname{stream-index}},["(", arguments, ")"];$
                                               & $lendStream_1$                                   \\
$\ebnfname{request-event} = name,["[",arguments,"]"];$
                                               & $abort~ask[\bar x_i]$                            \\
$\ebnfname{answer-value} = \ebnfname{stream-value} \mid \ebnfname{stream-failed} \mid \ebnfname{stream-complete};$
                                               & $v_i~err~done$                                   \\
$\ebnfname{answer-event} = \ebnfname{stream-variable},":","=",\ebnfname{answer-value};$
                                               & $\bar x_i := v_i$                                \\
$\ebnfname{port-index}  = [\ebnfname{number} \mid \ebnfname{variable}];$
                                               & $s$\\
$\ebnfname{port} = (\ebnfname{uppercase-letter}, \{\ebnfname{uppercase-letter}\})_{\ebnfname{port-index}};$
                                               & $SSU_1~DI$\\
$\begin{aligned}
\ebnfname{event} = [ port, ":" ],(& \ebnfname{method-call-event} \mid \ebnfname{request-event} \mid \ebnfname{answer-event} \\
& \mid \ebnfname{empty-event});
\end{aligned}$ 
                                               & I: abort\\
$\ebnfname{history} = event, \{",", event\};$ 
                                               & $ask[\bar x_1],\bar x_1 := v_1$                                          \\
\end{tabular}
    \caption[Syntax of events.]{Syntax of events in EBNF (also using overline, underline, and subscript syntax).}
\label{tb:syntax-events}
\end{table}

Sometimes multiple histories may be possible for a given protocol or module execution. We capture the underlying structure with a \textit{partial order on events}, as illustrated in Table~\ref{tb:syntax-partial-order}. The temporal dependency between events is represented with the $\rightarrow$ operator. Concurrent events are represented with the $\wedge$ operator.  A choice among multiple mutually-exclusive choices is represented with the $\mid$ operator. We use partial orders to describe a protocol as a sequence of events that captures all possible inter-leavings and possibilities concisely.

\begin{table}[htbp]
\centering
\begin{tabular}{ll}
\textbf{Syntax} & \textbf{Examples} \\
$\begin{aligned}
\ebnfname{partial-order} & = event\\
                         & \mid \ebnfname{partial-order}, "\rightarrow", \ebnfname{partial-order}, \{"\rightarrow", \ebnfname{partial-order} \} \\
                         & \mid \ebnfname{partial-order}, "\wedge", \ebnfname{partial-order} \\
                         & \mid \ebnfname{partial-order}, "\mid", \ebnfname{partial-order} \\
                         & \mid "(", \ebnfname{partial-order}, ")";\\
\end{aligned}$ & 
$\begin{aligned}
& abort \\
& ask[\bar x_1] \rightarrow \bar x_1 := v_1 \\
& \bar x_1 := v_1 \wedge \bar x_1 := v_2 \\
& abort[\bar x_1] \mid error[err, \bar x_1] \\
& (abort) \\
\end{aligned}$\\
\end{tabular}
    \caption[Syntax of partial orders.]{Syntax of partial orders in Extended Backus-Naur form with examples. Operators in order of priority: $\rightarrow$, $\wedge$, $\mid$. $\rightarrow$, $\wedge$, $\mid$ are all associative.}
\label{tb:syntax-partial-order}
\end{table}

Pull-stream modules react to external events by initiating new events. The syntax for describing the rules that describe their behavior is given in Table~\ref{tb:syntax-rules}. It is built around an \textit{antecedent} that is itself a partial order augmented with additional operations. \textit{Relations} are used to reason about stream variables and values and decide whether a rule applies for a given history. The rest of the antecedent operators are essentially a first-order logic with conjunction $\wedge$ (which also has the meaning of concurrent), disjunction $\mid$ (which also has the meaning of mutual exclusion), negation $\neg$, and \textit{quantifiers}, to reason about all possible events of a certain type, or finding events in a history that satisfy some conditions. A rule $\Rightarrow$ is a combination of an antecedent and the event it generates if the antecedent is true.

\begin{table}[htbp]
\centering
\begin{tabular}{ll}
\textbf{Syntax}                                & \textbf{Examples}                                \\
$\ebnfname{quantifier} = "\exists"_{\{\ebnfname{variable}\}} \mid "\forall"_{\{\ebnfname{variable}\}};$ 
                                               & $\exists_i$\\
$\ebnfname{relation-operator} = "=" \mid "\neq" \mid  "<" \mid  "\leq" \mid  ">" \mid  "\geq";$
                                               & $<~\geq$ \\
$\ebnfname{relation-operand}  = \ebnfname{stream-index-variable} \mid variable \mid number;$
                                               & $i~n~1$ \\
$\begin{aligned}
\ebnfname{relation} = &~ \ebnfname{relation-operand}, \ebnfname{relation-operator}, \ebnfname{relation-operand},\\
& \{\ebnfname{relation-operator}, \ebnfname{relation-operand}\};
\end{aligned}$
                                               & $r < i < n$ \\
$\begin{aligned}
\ebnfname{antecedent} & = boolean \mid event \mid relation \\
                      & \mid antecedent, "\rightarrow", antecedent, \{"\rightarrow", antecedent\} \\
                      & \mid antecedent, "\wedge", antecedent \\
                      & \mid antecedent, "\mid", antecedent \\
                      & \mid "(", antecedent, ")" \\
                      & \mid "\neg", antecedent \\
                      & \mid quantifier, antecedent; \\
\end{aligned}$
                                               & $i<n \wedge I: ask[\bar x_i]$ \\
$\begin{aligned}
\ebnfname{rule} & = antecedent, "\Rightarrow", event\\
                & \mid event, "\Leftarrow", antecedent;\\
\end{aligned}$                                 & $C: ping[\bar x_i] \Rightarrow S: \bar x_i := v_i$ \\
\end{tabular}
    \caption[Syntax of rules.]{Syntax of rules in Extended Backus-Naur form (also using overline, underline, subscript, and superscript syntax) with examples. Operators in order of priority: \textit{relational-operator}, $\rightarrow$, $\neg$, \textit{quantifier}, $\wedge$, $\mid$.  $\rightarrow$, $\wedge$, $\mid$ are all associative.}
\label{tb:syntax-rules}
\end{table}

To conclude the syntax, when multiple choices are possible for events in a partial order or an antecedent, it may be hard to read and parse them. We therefore sometimes use a vertical choice operator as syntactic sugar, shown in Figure~\ref{fig:syntactic-sugar}.

\begin{figure}[htbp]
\centering
\[
\begin{aligned}
\begin{sqcases}
expr_1\\
expr_2\\
...\\
expr_n\\
\end{sqcases}       & = "(",expr_1,"\mid",expr_2,"\mid", ...,"\mid", expr_n,")"
\end{aligned}
\]
\caption[Syntactic sugar.]{Syntactic sugar for a disjunction of antecedents or partial orders.}
\label{fig:syntactic-sugar}
\end{figure}

\subsection{Semantics}
\label{Section:Semantics}

We explain the semantics for the syntax described previously in this section. We simplify its description by first normalizing antecedents and partial orders with rewrite rules. We then describe what a history is and provide operations on it and how it is extended. We finish the section with a simple ping-pong protocol to illustrate how the semantics work.  

In the next sections, we do not describe the meaning of a partial order because it is the same as an antecedent. Any rules for antecedent that use the same syntax therefore also apply to partial orders. For conciseness, we use the following aliases for booleans, relations, antecedents, and events:
\[
\begin{aligned}
& b ::= boolean, \quad r ::= relation, \quad a ::= antecedent, \quad e ::= event~\text{except}~empty\\
\end{aligned}
\]

\subsubsection{Normalization}
\label{Section:Normalization}

We normalize partial orders and antecedents to make it easier to reason about them. The process consists in rewriting the expressions such that there are only sequences of events separated by conjunctions ($\wedge$) and disjunctions ($\mid$). A sequence is an arbitrarily large but finite list of events that happen before one another defined as:

\[
\begin{aligned}
& seq ::= \begin{cases}
  e \\
  seq \rightarrow e
\end{cases} \\
\end{aligned}
\]
 
The rewriting rules are shown in Figure~\ref{fig:normalization}. The first rule says that a sequence $seq_1$ followed by two concurrent sequences $seq_2$ and $seq_3$ is the same as the conjunction of $seq_1$ followed by each of the sequences. The second rule is similar with two concurrent sequences followed by a single sequence. The third rule says that a sequence $seq_1$ followed by a choice of either two sequences $seq_2$ or $seq_3$ is the same as a choice between the sequence $seq_1$ immediately followed by $seq_2$ or $seq_1$ immediately followed by $seq_3$. All the other rewriting rules show how to remove an empty event from a sequence or an antecedent.

\begin{figure}
\[
\begin{aligned}
seq_1 \rightarrow (seq_2 \wedge seq_3) & \leadsto (seq_1 \rightarrow seq_2) \wedge (seq_1 \rightarrow seq_3)\\
(seq_2 \wedge seq_3) \rightarrow seq_1 & \leadsto (seq_2 \rightarrow seq_1) \wedge (seq_3 \rightarrow seq_1) \\
seq_1 \rightarrow \begin{sqcases}
seq_2\\
seq_3
\end{sqcases} \leadsto seq_1 \rightarrow (seq_2 \mid seq_3) & \leadsto (seq_1 \rightarrow seq_2) \mid (seq_1 \rightarrow seq_3)\\
\begin{rsqcases}
seq_1\\
seq_2
\end{rsqcases} \rightarrow  seq_3 \leadsto (seq_1 \mid seq_2) \rightarrow  seq_3 & \leadsto (seq_1 \rightarrow seq_3) \mid (seq_2 \rightarrow seq_3)\\
seq \rightarrow empty \leadsto seq & \qquad \qquad empty \wedge a \leadsto a\\
empty \rightarrow seq \leadsto seq & \qquad \qquad a \wedge empty \leadsto a\\
seq_1 \rightarrow empty \rightarrow seq_2 \leadsto seq_1 \rightarrow seq_2 & \qquad \qquad a \mid empty \leadsto a \\
& \qquad \qquad empty \mid a \leadsto a\\
\end{aligned}
\]
\caption[Rewriting rules for normalization.]{Rewriting rules for normalization. The left-hand side of the $\leadsto$ operator is rewritten to its right-hand side.}
\label{fig:normalization}
\end{figure}

We have verified that the normalization terminates by using the Aprove method~\cite{giesl2017analyzing}, by testing the rules listed in Appendix~\ref{Section:Normalization-Rules} with an automated assistant~\cite{aprove-assistant}. This ensures that all possible antecedent or partial order can be rewritten in a finite number of steps.

\subsubsection{History}

The following definitions define what a \textit{history} is and various operations on it.

A history $H$ is a stream of past events that is either empty or has an event following a shorter history:
\[
H ::= \begin{cases}
\langle\rangle\\
H,e
\end{cases}
\]

The definition of a history is similar to a sequence of events $seq$. A history may contain additional events that are not part of a defined sequence while still correctly following that sequence.

Similar to the syntax definition in Figure~\ref{tb:syntax-events}, an event is either a function call, a request, or an answer (there are no empty event in a history). 

Every event of a history $H$ is syntactically unique. There is no variable in the syntax of events in a history, every \textit{stream-index}  and \textit{port-index} is a number and not a variable. Therefore we define two events as equal if they are written with exactly the same symbols in the same positions.

Consequently, an event $e$ is part of a history $H$, written $e \in H$, if an event that appears in $H$ is equal to $e$. The opposite relation is the negation of the previous: 

\[
\begin{aligned}
e \in H ::= & \begin{cases}
H=\langle\rangle & false\\
H=(H',e') & \begin{cases}
e=e' &  true \qquad \qquad \qquad e \notin H ::= \neg (e \in H)\\
e\neq e' & e \in H'
\end{cases}
\end{cases}
\end{aligned}
\]

The \textit{depth} of an event, written $depth(e,H)$, represents how far it appeared in the past. It is determined by counting the number of events that happened after until now. The operation is only defined if $e \in H$:

\[
\begin{aligned}
depth(e,H) ::= &~
H=H',e';~ \begin{cases}
e=e' &  0\\
e\neq e' & 1+depth(e,H')
\end{cases}
\end{aligned}
\]

An event $e_2$ therefore \textit{follows} an event $e_1$ if they are both part of the same history and $e_1$ is further in the past (i.e. its depth is greater). The $\textit{before}_H$ relationship is not defined if one or both of the operands are not in the history:
\[
\begin{aligned}
e_1~\textit{before}_H~e_2 & ::= e_1 \in H \wedge e_2 \in H  \wedge (depth(e_1,H) > depth(e_2,H))
\end{aligned}
\]

An event $e_2$ \textit{always follows} an event $e_1$ if the relationship is true for all possible histories. 

History $H$ \textit{entails} ($\vDash$) an expression $a$ (antecedent), which can be a complex expression that represents events, boolean values, relation operations between integers, or logic operations on expressions, if the following inference rules are \textit{true}\footnote{The bottom of an inference rule is true if all the conditions on top are true.}:
\[
\begin{aligned}
& \frac{e \in H}{H \vDash e} \quad
\frac{seq = e \quad e~\textit{before}{H}~e'}{H \vDash seq \rightarrow e'} \quad
\frac{seq = seq' \rightarrow e \quad H \vDash seq' \quad e~\textit{before}_{H}~e'}{H \vDash seq \rightarrow e'} \quad
\frac{b = \top}{H \vDash b} \quad
\\
& 
\frac{\frac{r~is~true}{r}}{H \vDash r} \quad
\frac{a = r \quad H \vDash r}{H \vDash a} \quad 
\frac{a = e \quad H \vDash e}{H \vDash a} \quad 
\frac{a = seq \quad H \vDash seq}{H \vDash a} \quad 
\frac{a = b \quad H \vDash b}{H \vDash a} \quad 
\frac{ \neg H \vDash a}{H \vDash \neg a} \quad 
\\
&
\frac{H \vDash a_1 \quad \neg H \vDash a_2}{H \vDash a_1 \mid a_2} \quad
\frac{\neg H \vDash a_1 \quad H \vDash a_2}{H \vDash a_1 \mid a_2} \quad
\frac{H \vDash a_1 \quad H \vDash a_2}{H \vDash a_1 \wedge a_2} \quad
\frac{ \exists_{vars} H \vDash a}{H \vDash \exists_{vars} a} \quad
\frac{ \forall_{vars} H \vDash a}{H \vDash \forall_{vars} a} \quad
\\
\end{aligned}
\]

By convention, indexes for stream variables and stream values start at $1$ and increase by one for each immediately following item. If an antecedent has free variables in it, i.e. variables not mentioned in a quantifier, it is assumed that the quantifier $\exists_{i \in \mathcal{N}}$ for each free variable $i$. For example, suppose a history $H = ask[\bar x_1], abort$. Then $H \vDash ask[\bar x_i]$ because $i$ is a free variable in $ask[\bar x_i]$, which is equivalent to writing $H \vDash \exists_{i \in \mathcal{N}} ask[\bar x_i]$. For $i=1$, $H \vDash ask[\bar x_1]$, because $ask[\bar x_1] \in H$ (it is the first element by definition). The use of stream index variables therefore enables matching \textit{series} of events in a history.

\subsubsection{History Progression}

The specification for modules uses a single rule that defines the behavior of the implication $a \Rightarrow e$. Intuitively, if the antecedent expression $a$ is entailed by the the current history $H \vDash a$ and the event $e$ has not already happened\footnote{Remember that each event is syntactically unique.}, then $e$ should eventually happen and the new history $H'$ will be the old history $H$ extended by $e$. During the execution, it is possible that multiple implication rules may match at the same time, which enables concurrency. In this case, an implementation is free to execute them in any order. To be correct, an implementation needs to correctly follow the pull-stream protocol regardless of which rule was executed first.

Formally, we therefore define the behavior of the implication ($\Rightarrow$) as:
\[
\begin{aligned}
& H.(a \Rightarrow e) \succ H' \quad
\frac{H \vDash a \quad e \notin H}{H. (a \Rightarrow e) \succ H,e} \quad
\frac{\neg H \vDash a}{H. (a \Rightarrow e) \succ H} \quad
\frac{e \in H}{H. (a \Rightarrow e) \succ H}\\
\end{aligned}
\]

Finally, note that by definition, $a \Rightarrow e$ implies that for any event $e'$ in $a$, $H \vDash e' \rightarrow e$. As a final remark, it is interesting to note that a history is itself a stream of events and can therefore be represented and processed using pull-streams!

\subsubsection{Step-by-Step Ping Pong Example}

We illustrate the syntax and semantics with a simple ping-pong protocol based on a single request event $ping[\bar x_i]$ and a single answer event $\bar x_i := pong$\footnote{In this protocol, $\forall_i v_i = pong$.}. Let's assume a client $C$ initiates a ping and a server $S$ answers with a pong. The client and the server are respectively ports on which events happen. 

The client sends a first ping, waits for an answer from the server and then sends the next ping, infinitely often. The behavior of the client can be described with the following rules:
\[
\begin{aligned}
& C: ping[\bar x_1] \Leftarrow \neg C: ping[\bar x_1]\\
& C: ping[\bar x_i] \Leftarrow S: \bar x_{i-1} := pong
\end{aligned}
\]
The server answers each ping with a pong infinitely often. Its behavior can be described with the following rule:
\[
S: \bar x_i := pong \Leftarrow C: ping[\bar x_i]
\]

As you have seen, by convention we write the event initiated by the port on the left and the events on which it depends on the right. An execution of that protocol according to the semantics would follow those steps. Initially, the history is empty $(H=\langle\rangle$). The only rule that can extend the history is the first client rule and since there are no event in the history, it is true that $C: ping[\bar x_1] \notin H$. All other rules are missing a past event on which they depend and therefore cannot extend the history.

The first rule therefore applies, and now $H = \langle\rangle, C: ping[\bar x_1]$. The first rule of $C$ no longer applies (and will never again). The second rule of $C$ still needs an answer. The only rule that applies is that of $S$. $H$ is again extended and is now $H = \langle\rangle, C: ping[\bar x_1], S: \bar x_1 := pong$. The rule of $S$ no longer applies. But the second rule of $C$ now does, because $\exists_{i \in \mathcal{N}}$ when $i=2$. In such a case, the rule is equivalent to:
\[
C: ping[\bar x_2] \Leftarrow S: \bar x_1 := pong\\
\]

It applies because $S: \bar x_1 := pong \in H$. Then $H$ is extended with the new ping, and so on and so forth forever. The invariant of the ping-pong protocol may be abstracted as the following partial-order:
\[
C: ping[\bar x_i] \rightarrow S: \bar x_i := pong\\
\]

And a corresponding generator function that creates sequences of events according to the protocol for a finite number of steps can be written:

\begin{figure}[htbp]
\begin{flushleft}
\textbf{Signature}: $ping\_pong\_sequence(\underline{n},C,S)$\\
\textbf{Parameters}:\\
$\underline{n}$ is the number of ping requests.\\
\textbf{Implementation}:\\
\end{flushleft}
\[
\begin{aligned}
& \underline{events}(\underline{i}) ::= \begin{cases} 
      i \leq n   & C: ping[\bar x_i] \rightarrow S: \bar x_i := pong \rightarrow \underline{events}(i+1)) \\
      i > n      & \textit{empty} \\
\end{cases} \\
& \textbf{returns}~\underline{events}(1)
\end{aligned}
\]
\caption[Ping-pong partial order.]{\label{fig:pingpong} Procedure to generate a partial order of events for the ping-pong protocol.}
\end{figure}

We now use a similar exposition to present the pull-stream protocol, first by introducing the protocol and then reference modules that generate valid sequences of events.

%% file: protocol.tex
In this section, we define what sequences of events follow the pull-stream protocol at a given interface between two modules. There was no previous formal specification that described it, we obtained them by asking questions to the community about the usual expectations on the behavior of pull-stream modules\footnote{The entire discussion is in this issue on the main pull-stream repository on GitHub: \url{https://github.com/pull-stream/pull-stream/issues/100}.} and by testing existing modules. We reformulated them to be concise, precise, and complete. 

\subsection{Overview}

The protocol covers two possible sequences, a \textit{normal sequence} in which the upstream module produces all its values and then stops, and an \textit{early termination sequence} in which the upstream module is terminated by the downstream module before all values have been produced. We cover both in turn. Note that the protocol only specifies the sequence of events between two modules on a single interface: it does not specify constraints \textit{between} two interfaces, such as between the input and output of a transformer. These constraints are instead part of the \textit{specification} of a module and therefore allow module designers a maximum amount of flexibility.

To present both the normal sequence and the early termination sequence, we start with concrete sequences of events. Since many sequences are similar to one another apart from the specific ordering of events, we then capture the regularity in a partial order using the \textit{happens before} ($e_1 \rightarrow e_2$) relation between events. Finally, we generalize the partial orders to functions that generates them for specific cases given some parameters that are specific to each of the two cases. These functions are described in Figure~\ref{fig:Pull-Stream-Protocol-Normal-Sequence} and \ref{fig:Pull-Stream-Protocol-Early-Termination-Sequence}. 

Note that the current protocol mostly forbids concurrent requests\footnote{The early termination is an exception to the rule to allow aborting before an answer has been provided.} or out-of-order answers. These additional cases are not supported to simplify the implementation of modules in JavaScript. We discuss the additional possibilities that could be offered by the protocol in Appendix~\ref{Appendix:Concurrent-Variations}, which may be easier to use in languages with dataflow variables such as Oz. These additional cases illustrate the expressiveness of our notation for concurrency. Readers solely interested in the pull-stream protocol implementation for JavaScript may safely skip these explanations.

\subsection{Pull-Stream Events}
Table~\ref{tb:js-events} showed examples of the events of the pull-stream protocol. Figure~\ref{fig:Pull-Stream-Events} present them again regrouped in different categories to ease the presentation of the protocol and later specifications.

\begin{figure}[htbp]
    \centering
    \begin{subfigure}[t]{0.48\textwidth}
        \begin{align*}
\underline{terminate}(\bar x_i) & ::= 
\begin{sqcases}
abort[\bar x_i]\\
error[err, \bar x_i]
\end{sqcases}\\ 
\underline{request}(\bar x_i)      & ::= 
\begin{sqcases}
ask[\bar x_i]\\
\underline{terminate}(\bar x_i)\\
\end{sqcases}
\end{align*}
        \caption{\label{fig:requests} Requests.}
    \end{subfigure}
    ~ 
    \begin{subfigure}[t]{0.48\textwidth}
\begin{align*}
\underline{terminated}(\bar x_i) & ::= 
\begin{sqcases}
\bar x_i := done \\
\bar x_i := err
\end{sqcases}\\
\underline{answer}(\bar x_i) & ::= 
\begin{sqcases}
\bar x_i := v_i\\
\underline{terminated}(\bar x_i)\\
\end{sqcases}
\end{align*}        
         \caption{\label{fig:answers}  Answers.}
    \end{subfigure}
    \caption[Pull-stream protocol events.]{\label{fig:Pull-Stream-Events} All possible pull-stream events regrouped by category under the $\underline{\textit{request}}, \underline{\textit{terminate}}, \underline{\textit{answer}},$ and $\underline{\textit{terminated}}$ functions.}
\end{figure}

All the \textit{requests} are initiated downstream. All requests expect an answer in $\bar x_i$ ($request(\bar x_i)$). \textit{Terminate} requests are used to terminate the stream \textit{early} before all values have been produced. A normal termination, such as when no more values are needed, is performed with $abort[\bar x_i]$. An abnormal termination, such as when an internal error prevents a module from receiving more values, is performed with $error[err, \bar x_i]$. All \textit{answers} are initiated upstream and provide either a \textit{value} $v_i$ or signal that the stream has \textit{terminated} either normally with $done$ or abnormally with $err$.

\subsection{Normal Sequence}

The \textit{normal sequence} represents sequences of events in which the number of \textit{ask} requests ($r$) from the module downstream is strictly greater than the number of values produced ($n$). Therefore the upstream module produces all its values and then terminates the stream. After receiving a terminated answer, the module downstream \textit{must} never make additional requests, therefore  $r=n+1$.

We describe the interaction between the input $I$ of a downstream module and the output $O$ of the module immediately upstream. For example, in the case where two requests are made in sequence ($I: ask[\bar x_1]$ and $I: ask[\bar x_2]$), and one value ($O: \bar x_1 := v_1$) and a terminated event ($O: terminated(\bar x_1)$) are produced, there are all six possible sequences of events.

The first two cases correspond to \textit{co-routining} between the downstream module and the upstream module: the downstream module \textit{asks} for a value, waits for an answer, then \textit{asks} the next value. 
\[
\begin{aligned}
\begin{aligned}
I: ask[\bar x_1], \\
I: ask[\bar x_1], \\
\end{aligned}
&&
\begin{aligned}
& O: \bar x_1 := v_1, \\
& O: \bar x_1 := v_1, \\
\end{aligned}
&&
\begin{aligned}
& I: ask[\bar x_2], \\
& I: ask[\bar x_2], \\
\end{aligned}
&&
\begin{aligned}
& O: \bar x_2 := done \\
& O: \bar x_2 := err \\
\end{aligned}
\end{aligned}
\]

We abstract the two termination cases ($\bar x_2 := done$ and $\bar x_2 := err$) in a single abstract event $terminated(\bar x_2)$ and express the ordering constraint as a sequence\footnote{As explained in Section~\ref{Section:Normalization} using a sequence rather than a concrete history allows multiple concurrent streams to generate their events in a single global history of interleaved events.}: 
\[
\begin{aligned}
I: ask[\bar x_1] \rightarrow O: \bar x_1 := v_1 \rightarrow I: ask[\bar x_2] \rightarrow O: terminated(\bar x_2) \\
\end{aligned}
\]

The four other cases correspond to having two \textit{concurrent} $asks$ made before the answers have arrived, two cases in which the answers arrive in order and two out-of-order. These last four cases could possibly improve performance in some cases but implementations of pull-stream modules do not use them because the added implementation complexity is not worth the gain. They are not necessary to understand the protocol as it is currently used in practice but the interested reader may still find a complete description in Appendix~\ref{Section:Concurrent-Normal-Sequence}. 

We generalize the \textit{co-routining} sequence to an arbitrary number of values $n$ and any pair of input and output ports $I$ and $O$ with the \textit{normal\_sequence} function of Figure~\ref{fig:Pull-Stream-Protocol-Normal-Sequence}. The function should only be used if $r=n+1$, otherwise the other function of Figure~\ref{fig:Pull-Stream-Protocol-Early-Termination-Sequence} should be used\footnote{$r > n+1$ is incorrect.}. The \textit{normal\_sequence} function generates a sequence one request and answer at a time starting from the first ($events(1)$). There are two different cases for generating a new request and answer. If $i \leq n$ than a request should be answered by a value before the next request is initiated. If $i = n+1$ then there are no more values and the request will be answered by a \textit{terminated} event.

\begin{figure*}[htbp]
\begin{flushleft}
\textbf{Signature}: $normal\_sequence(n,I,O)$\\
\textbf{Parameters}:\\
$n$ is the number of values in the stream $(n \geq 0)$\\
\textbf{Implementation}:\\
\end{flushleft}
\[
\begin{aligned}
& \underline{events}(\underline{i}) ::= \begin{cases} 
      i \leq n   & I: ask[\bar x_i] \rightarrow O: \bar x_i := v_i \rightarrow\underline{events}(i+1)) \\
      i = n+1   & I: ask[\bar x_i] \rightarrow O: \underline{terminated}(\bar x_i)
\end{cases} \\
& \textbf{returns}~\underline{events}(1)
\end{aligned}
\]
\caption[Normal Sequence.]{\label{fig:Pull-Stream-Protocol-Normal-Sequence} Procedure to generate a partial order of events for the normal sequence (in this case, the partial order is a sequence).}
\end{figure*}

The function can be used to generate all possible sequences of events. For example, for $n=1$ the partial order of events for the normal sequence generated by the function is:
\[
\begin{aligned}
I: ask[\bar x_1] \rightarrow O: \bar x_1 := v_1 \rightarrow I: ask[\bar x_2] \rightarrow O: \underline{terminated}(\bar x_2) \rightarrow empty\\
\end{aligned}
\]

Making both $terminated(\bar x_i)$ cases explicit:
\[
\begin{aligned}
I: ask[\bar x_1] \rightarrow O: \bar x_1 := v_1 \rightarrow I: ask[\bar x_2] \rightarrow 
\begin{sqcases}
O: \bar x_2 := done\\
O: \bar x_2 := err \\
\end{sqcases}
\end{aligned}
\]

Using the rewriting rules of Section~\ref{Section:Normalization} we obtain the two possibilities that correspond to the two examples given at the beginning of the section:
\[
\begin{aligned}
& (I: ask[\bar x_1] \rightarrow O: \bar x_1 := v_1 \rightarrow I: ask[\bar x_2] \rightarrow O: \bar x_2 := done) \mid \\
& (I: ask[\bar x_1] \rightarrow O: \bar x_1 := v_1 \rightarrow I: ask[\bar x_2] \rightarrow O: \bar x_2 := err)
\end{aligned}
\]

\subsection{Early-Terminated Sequence}

The \textit{early\_terminated\_sequence} represents sequences of events in which the number of $ask$ requests is less or equal to the number of values that could be produced by the upstream module ($r \leq n$) . The last $ask$ request is always followed by a \textit{terminate} request. The upstream module is therefore \textit{terminated} earlier than in the normal sequence and the stream is potentially shorter than it would have been otherwise\footnote{For the case where $r=n$ the upstream module actually terminate on the same request index it would have in the normal sequence. However, it terminates on a $terminate$ request rather than an $ask$ request.}.

For example, in the case of a stream of two values ($n=2$), terminated early on the second request ($r=1$), the following sequences of events are possible. In these sequences, we do not list the normal and abnormal termination requests and terminated answers explicitly, we use $I: \underline{terminate}(\bar x_i)$ and $O: \underline{terminated}(\bar x_i)$ instead.

The \textit{co-routining} case is analogous to the \textit{normal sequence}: an answer is expected before the next request is initiated, but the last request is a termination request instead of an $ask$ request and the second value is ignored: 
\[
\begin{aligned}
& I: ask[\bar x_1],~ O: \bar x_1 := v_1,~I: \underline{terminate}(\bar x_2),~O:\underline{terminated}(\bar x_2)\\
\end{aligned}
\]

At first glance, it may seem that it is sufficient to support early termination. However, because an answer is expected for each request before issuing the next, it is not possible to abort the upstream module if an answer takes too long to arrive. Because of that limitation, a limited form of \textit{in-order} concurrency is supported: a terminate request may be concurrently initiated if the last answer takes too long to arrive. The previous answer is terminated as soon as possible, and the answer for the termination request comes right after:
\[
\begin{aligned}
& I: ask[\bar x_1],~I: \underline{terminate}(\bar x_2),~O: \underline{terminated}(\bar x_1)~O: \underline{terminated}(\bar x_2)\\
\end{aligned}
\]

As for the normal sequence, there are other possible concurrent cases that are not used. The interested reader may still find a complete description in Appendix~\ref{Section:Concurrent-Early-Termination-Sequence}.

As for the normal sequence, we generalize the sequence to an arbitrary number of values $n$, number of  $ask$ requests $r$, with an additional parameter $w$ is used to distinguish between the case where the last answer was waited for or not. The generalization for the \textit{early\_terminated\_sequence} function is shown in Figure~\ref{fig:Pull-Stream-Protocol-Early-Termination-Sequence}. This function applies in cases where $r \leq n$, otherwise the \textit{normal\_sequence} applies.

 The function generates a partial order, one request and answer at a time starting from the first ($events(1,empty)$). Note that it uses $T$ as a parameter to save the terminated answer in case $w=\top$ to place it correctly later in the sequence. There are four different cases in the generation of events. If $i \leq r-1$ then a value matching the $ask$ is produced. If $i=r$ and the answer is waited for ($w=\top$), it is similar to the first case. If $i=r$ but the answer is not waited for ($w=\top$), then an $ask$ with a terminated answer is produced and saved for later in $T$. In the last case where $i=r+1$, the terminate request is generated and an additional terminated answer $T$, empty or not depending on whether $w$ was $\top$ or $\bot$, and finally the answer for the terminate request.

\begin{figure*}[htbp]
\begin{flushleft}
\textbf{Signature}: $early\_terminated\_sequence(n,r,w,I,O)$\\
\textbf{Parameters}:\\
$n$ is the number of values in the stream $(n \geq 0)$\\
$r$ is the number of $ask$ requests performed from downstream before terminating $(0 \leq r \leq n)$\\
$w$ is whether the last value had been waited for before terminating ($w=\top$) or not ($w=\bot$)\\
\textbf{Implementation}:\\
\end{flushleft}
\[
\begin{aligned}
& \underline{events}(\underline{i},\underline{T}) ::= \begin{cases} 
      i \leq r-1          & I: ask[\bar x_i] \rightarrow O: \bar x_i := v_i \rightarrow \underline{events}(i+1,T) \\
      i = r \wedge w      & I: ask[\bar x_i] \rightarrow O: \bar x_i := v_i \rightarrow \underline{events}(i+1,T) \\
      i = r \wedge \neg w & I: ask[\bar x_i] \rightarrow \underline{events}(i+1, O: \underline{terminated}(\bar x_i)) \\
      i=r+1               & I: \underline{terminate}(\bar x_i) \rightarrow T \rightarrow O: \underline{terminated}(\bar x_i)
\end{cases} \\
& \textbf{returns}~\underline{events}(1,empty)
\end{aligned}
\]
\caption[Early-Termination Sequence]{\label{fig:Pull-Stream-Protocol-Early-Termination-Sequence} Procedure to generate a partial order of events for early terminated sequences.}
\end{figure*}

\subsection{Correctness}

The sequence of events (\textit{history} $H$) observed at the interface of two modules with input ($I$) and output ($O$) is \textit{correct} if it follows either case of the pull-stream protocol and does not generate more events than those. In practice, we check the conformity of actual executions with the following invariants implemented in a transformer module~\cite{pull-stream-protocol-checker} placed between two other modules:
\begin{enumerate}
	\item No additional request after a \textit{terminate} request or a \textit{terminated} answer;
    \item Every expected answer ($\bar x_i$) eventually happens; 
    \item Every expected answer ($\bar x_i$) happens only once;
    \item Expected answers happen in the creation order of their stream variable ($\bar x_i$);
    \item No concurrent $ask$ requests;
    \item (If the stream is finite), the stream is eventually terminated.
\end{enumerate}

%% file: model-modules.tex
In this section, we provide reference modules that can guide a correct and complete implementation. They are useful both as an illustration of our notation for specifying the behavior of modules as well as for testing other modules. For the latter case, we implemented them in JavaScript~\cite{pull-stream-reference-modules}\footnote{The specifications given in this section follow the pull-stream protocol presented in the last section. The JavaScript implementations might have additional parameters compared to the specifications to change the order of event execution and intentionally generate incorrect behaviors to see how other modules react to them. They may also use different conventions that are more idiomatic to JavaScript or that follow the existing naming conventions of the pull-stream community.} and we report on our experiments in testing community modules with them in Section~\ref{Section:Evaluation}.

\begin{figure}[htbp]
    \begin{center}
    \includegraphics[width=0.5\textwidth]{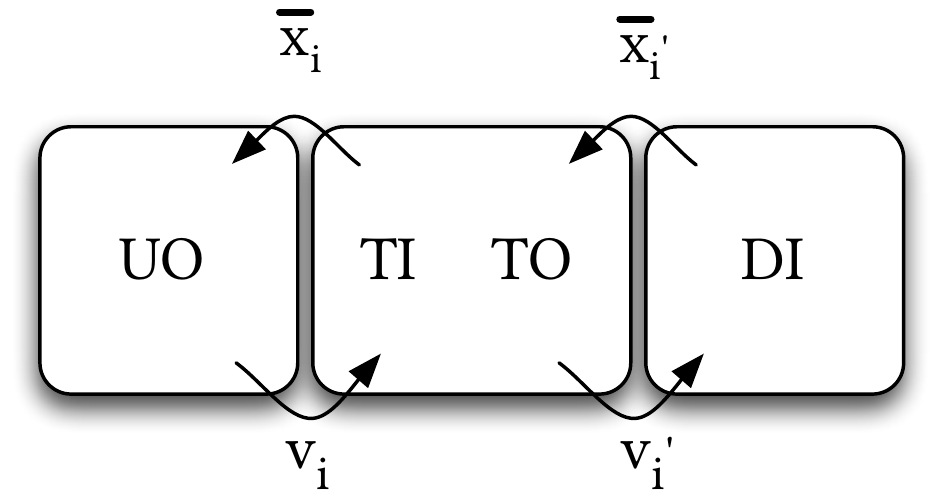}
    \end{center}
    \caption[Abstract representation of modules.]{\label{fig:ModelModules} Abstract representation of the interfaces between a transformer and the output of the module upstream (UO-TI), which could be a source or another transformer, and the input of the module downstream (TO-DI), which could be a sink or another transformer. Each interface generates a variable stream with a unique name (ex: $\bar x_i$). Each new variable is implicitly added by a request coming from the interface's downstream port (ex: $TI$). The answers to those requests produce corresponding values that flow in the opposite direction (ex: $v_i$ and $v_i'$).}
\end{figure}

To present the next modules, we use some conventions to make each interface between pairs of modules unique and provide some intuitions about how values flow through them. Figure~\ref{fig:ModelModules} shows the interfaces between a transformer module and the output of the module immediately upstream (UO) and the input of the module immediately downstream (DI). The module upstream may be a source or a transformer and the module downstream may be a sink or a transformer also. The interface upstream (UO-TI) receives requests which create an implicit stream of $\bar x_i$ variables and produces answers of values $v_i$. To show the progression of values through the modules, we write $v_i'$ the value in a downstream interface with the same index, after some processing has occurred. Correspondingly, the variable in the request for that answer is noted $\bar x_i'$.

A \textit{port} is used to specify the particular side of an interface where an event is initiated. When the events are implemented with function calls as in JavaScript, it represents the caller side. The destination of a request or an answer is implicit and not ambiguous because it is always the complementary port of the same interface (ex: $UO$ for a request initiated on $TI$). The rules that define the behavior of the modules are given according to the ports of the abstract representation given in Figure~\ref{fig:ModelModules}. In the rest of this section, the upstream module is a source with an output $UO$, the middle module is a transformer with an input $TI$ and an output $TO$, and the downstream module is a sink with an input $DI$. The three are connected in a single pipeline one immediately after the other ($UO-TI$ and $TO-DI$). We present all three modules in sequence and conclude with a discussion about completeness and correctness of specifications.


The reference source presented in Figure~\ref{fig:RulesSource} answers requests with either a stream value ($v_i$), a termination marker ($done$), or an error ($err$). The rules are parameterized by the number of stream values to produce ($n$) and a boolean flag ($err$) to simulate the behavior of modules that may fail to correctly produce a value instead of normally terminating.

\begin{figure}[htbp]
\begin{flushleft}
\textbf{Parameters:}\\
$n$ ($n \geq 0$): number of values to produce;\\
$err$ (\textit{boolean}): terminate with a done ($err=\bot$) or an error ($err=\top$).\\
\bigskip
\end{flushleft}
\[
\begin{aligned}
UO: \bar x_i & := v_i & \Leftarrow &~ TI: ask[\bar x_i] \wedge ~ i \leq n \\
UO: \bar x_i & := done       & \Leftarrow & \begin{sqcases}
TI: ask[\bar x_i]       \wedge ~ \neg err \wedge i = n+1 \\
TI: \underline{terminate}(\bar x_i) \\
\end{sqcases}\\
UO: \bar x_i & := err   & \Leftarrow &~ TI: ask[\bar x_i] \wedge ~ err \wedge i = n+1
\end{aligned}
\]
\caption{\label{fig:RulesSource} Rules for a reference source.}
\end{figure}

The behavior of the module is defined as implication rules ($e \Leftarrow a$) that are reversed to emphasize on the left side all the possible events initiated by the module. The antecedent $a$ is a conjunction of posssibly multiple things: (1) events initiated downstream from the transformer input ($TI$); (2) boolean relations between the stream index of a particular request ($i$) and one of the source parameters ($n$); (3) boolean flags ($err$). In the rules, $i$ in a $TI: request(\bar x_i)$ is a free variable that matches any event in the history regardless of its concrete index ($1,2,...$). If the antecedent is true according to the semantics (Section~\ref{Section:Semantics}) and the event $e$ in the consequent of the implication has not happened yet ($e \notin H$), then the implementation of the source should eventually initiate $e$. As an example for the first rule, if the transformer input has asked for a value in the past ($TI: ask[\bar x_i] \in H$), there are still values to produce ($i \leq n$), and $v_i$ has not been produced yet ($UO: \bar x_i := v_i \notin H$), then a value $v_i$ should eventually be produced and assigned to $\bar x_i$. The other rules are straight-forward to derive from the pull-stream protocol. The rules for the source have been designed to be mutually exclusive, therefore only one rule applies at a time\footnote{This is not true in general: some modules may coordinate multiple concurrent streams and therefore have many rules that apply at the same time.}.

The organization of the rules in the figure makes it easy to visually inspect that all possible answers have rules associated with them.  The same rules may also be rewritten in the other direction to verify that all cases of requests from the module immediately downstream are covered. Knowing that all requests and answers are both covered, we can be confident that the behavior of the module is fully specified according to the pull-stream protocol. The rules may be easily adapted to a different configuration. For example, if the source was connected directly to the sink, all rules with the prefix $TI$ could be replaced with the prefix $DI$ and would apply in the same way.

\begin{figure}[htb]
\begin{flushleft}
\textbf{Parameters:}\\
$r$ ($r \geq 0$): number of $ask$ requests to perform;\\
$err$ (\textit{boolean}): terminate with an abort ($err=\bot$) or with an error ($err=\top$);\\
$w$ (\textit{boolean}): wait for the previous value before terminating ($w=\top$) or not ($w=\bot$).\\
\end{flushleft}
\bigskip
\[
\begin{aligned}
\underline{wait}(j) ::= & \begin{sqcases}
w \wedge TO: \bar x_j' := v_j \\
\neg w \wedge DI: ask[\bar x_j']
\end{sqcases}\\
DI: ask[\bar x_1'] & \Leftarrow ~r > 0 \\
DI: ask[\bar x_i'] & \Leftarrow ~ i \leq r \wedge TO: \bar x_{i-1}' := v_{i-1}' \\
DI: abort[\bar x_1'] & \Leftarrow ~ r = 0 \wedge \neg err\\
DI: abort[\bar x_i'] & \Leftarrow ~ i = r+1 \wedge \neg err \wedge \underline{wait}(i-1) \\
DI: error[err, \bar x_1'] & \Leftarrow ~ r = 0 \wedge err\\
DI: error[err, \bar x_i'] & \Leftarrow ~ i = r+1 \wedge err \wedge \underline{wait}(i-1) \\
\end{aligned}
\]
\caption{\label{fig:RulesSink} Rules for a reference sink.}
\end{figure}

The reference sink in Figure~\ref{fig:RulesSink} makes at most $r$ $ask$ requests for values and an extra $terminate$ request if the source has not terminated yet. The sink may make fewer requests than there are values available. If fewer values are requested than are available ($r < n$), the sink terminates \textit{early}. Two other parameters control the termination behavior: (1) it may terminate normally ($err=\bot$) or abnormally ($err=\top$); (2) it may terminate after having received an answer ($w=\top$) or immediately after having issued an $ask$ request ($w=\bot$). If the upstream module returns a terminated answer in response to an \textit{ask} request rather than a value, the sink stops emitting events, as expected by the pull-stream protocol.

The rules are organized similarly as for the reference source. Since the event initiated by the sink are requests, they are put on the left side of the implication rules. The stream of requests is defined inductively. The initial request depends only on the module parameters. For example, the first request will be $DI: ask[\bar x_1']$ if there is at least one $ask$ request to perform. Subsequent requests, such as $DI: ask[\bar x_i']$, happen after an answer has been initiated from the transformer output (ex: $TO: \bar x_{i-1}' := v_{i-1}'$). In case the transformer module initiated a terminated answer, none of the rules apply and therefore the sink stops. Note that the rules use '$'$' to indicate the stream variables ($\bar x_i'$) and stream values ($v_i'$) follow one stage of processing that happens before the sink. If the sink was connected directly to the source with no transformer in-between, all '$'$' should be removed from the rules.

\begin{figure}[htbp]
\begin{flushleft}
\textbf{Parameters}:\\
$r$ ($r \geq 0$): number of $ask$ requests to perform; \\
$err$ (\textit{boolean}): terminate with an abort ($err=\bot$) or with an error ($err=\top$).\\
\end{flushleft}
\[
\begin{aligned}
TI: ask[\bar x_i] & \Leftarrow DI: ask[\bar x_i'] \wedge i \leq r \\
TI: abort[\bar x_i] & \Leftarrow 
\begin{sqcases}
DI: ask[\bar x_i'] \wedge i = r + 1 \wedge \neg err\\
DI: abort[\bar x_i'] \wedge \neg TI: \underline{terminate}(\bar x_{i-1})
\end{sqcases} \\
TI: error[err, \bar x_i] & \Leftarrow 
\begin{sqcases}
DI: ask[\bar x_i'] \wedge i = r + 1 \wedge err\\
DI: error[err, \bar x_i'] \wedge \neg TI: \underline{terminate}(\bar x_{i-1})
\end{sqcases} \\
TO: \bar x_{i}' := v_i' & \Leftarrow UO: \bar x_i := v_i\\
TO: \bar x_{i}' := done & \Leftarrow  
\begin{sqcases}
UO: \bar x_i := done\\
DI: \underline{terminate}(\bar x_i') \wedge TI: \underline{terminate}(\bar x_{i-1}) \wedge UO: \bar x_{i-1} := done
\end{sqcases}\\
TO: \bar x_{i}' := err & \Leftarrow 
\begin{sqcases}
UO: \bar x_i := err\\
DI: \underline{terminate}(\bar x_i') \wedge TI: \underline{terminate}(\bar x_{i-1}) \wedge UO: \bar x_{i-1} := err
\end{sqcases}
\end{aligned}
\]
\caption{\label{fig:RulesTransformer} Rules for a reference transformer.}
\end{figure}

The reference transformer in Figure~\ref{fig:RulesTransformer} propagates requests from its output to its input and similarly transforms the values received on its input and send them on its output. The parameters are similar to those of the sink: $r$ for the number of \textit{ask} requests and $err$ to control whether the termination is done normally or abnormally. It does not need a wait ($w$) parameter because its waiting behavior actually is the same as that of the sink downstream.

The rules presented are similar to the rules for the reference source and sink. They syntactically capture the transformation behavior by relating the non-transformed upstream stage ($\bar x_i, v_i$) and the transformed downstream stage ($\bar x_i', v_i'$). The rules are similar to those of the sink with a subtle addition to avoid terminating upstream more than once. The second case of $TI: abort[\bar x_i]$ and $TI: error[err, \bar x_i]$ ensure that if the transformer aborted early a second abort will not be issued. The second case of $TO: \bar x_i' := done$ and $TO: \bar x_i' := err$ ensure that the termination request from downstream will receive an answer after the answer upstream has been received, even if a request had not been issued to avoid terminating twice.

Our specification assumes the downstream module behaves correctly (ex: no additional $ask[\bar x_i']$ after an $abort$). If the downstream module were incorrect, it is possible that the incorrect behavior could be propagated upstream. In practice a protocol checker~\cite{pull-stream-protocol-checker} may be used to catch those incorrect behaviors without having to complicate the transformer specification to defensively handle incorrect behavior.

\subsection{Completeness and Correctness}

The specification of a module is \textit{complete} if it defines rules for all possible events that may happen externally. The specification is \textit{correct} (follows the pull-stream protocol) if when combined with one of the reference modules on all its interfaces, any possible history generated by their combined specification at each of these interfaces follows the pull-stream protocol. To test implementations, we use a JavaScript implementation of the reference modules~\cite{pull-stream-reference-modules} to generate all possible sequences of events on streams of a finite size and check that the events on all interfaces follow the pull-stream protocol with a protocol checker~\cite{pull-stream-protocol-checker}. We provide more details in the next section.

%% file: evaluation.tex
We tested our understanding of the protocol using the modules of the core library~\cite{pull-stream} and checked them for conformity at the same time. To do so, we generated all possible sequences of events using our references modules~\cite{pull-stream-reference-modules} and checked that the events generated at the interface of any two modules followed the pull-stream protocol with a checker module~\cite{pull-stream-protocol-checker}. Our experiments are available in a GitHub repository~\cite{ecoop18-pull-stream-experiments}. We have found that in almost all cases the implementations correctly follow the pull-stream protocol as described in this paper, which is a testament to the quality of the core library. However, we did find one inconsistency and one incorrect behavior according to our specification. 

The inconsistency was found in the way some source modules can be terminated without a callback but not others\footnote{Reported here \url{https://github.com/pull-stream/pull-stream/issues/101} and applicable to version 3.6.1.}. That behavior was part of our initial specification of the protocol but we since removed it after finding that many sources do not actually support it. 

The incorrect behavior was found in a corner case for the \textit{take} transformer module in which an early termination downstream triggers two abort events upstream\footnote{Reported here \url{https://github.com/pull-stream/pull-stream/issues/104}, same version.}. The erroneous behavior correspond to the behavior our reference transformer in Section~\ref{Section:Model-Modules} would have if the abort condition did not check for an existing termination request upstream (i.e. if the second case of $TI: abort[\bar x_i]$ were simply $DI: abort[\bar x_i']$). The expected behavior was not documented but following a discussion with Dominic Tarr\footnote{\textit{Idem}.}, it was confirmed it was incorrect. In practice, it seems most sources and transformers handle multiple termination requests just fine so that was not a major problem for interoperability. However, it did illustrate that even seemingly simple protocols can have surprising corner cases and that an effort at formalizing is beneficial to identify them.

%% file: related.tex
\textit{Pull-stream documentation.} The community that created and maintains pull-stream modules has also produced informal documentation about their expected behavior~\cite{pull-stream-specification,pull-stream-original-blog-post,pull-stream-examples}. Our description of expected sequences of events in the pull-stream protocol and our reference modules are consistent with that documentation and fill some gaps. To the best of our knowledge this paper is the first academic paper that describes the pattern formally and suggests a specification language for pull-stream modules. 

\textit{Streaming design patterns as libraries.} The pull-stream design pattern itself is similar to at least another open source JavaScript compositional library called Reducers written by Irakli Gozalishvili~\cite{reducers}, which was independently conceived around the same time. There are some other examples in academic publications of streaming libraries for other languages. Spark Streaming exists for Scala and was used for large-scale stream processing~\cite{zaharia2013discretized}. Biboudis et al.~\cite{biboudis2015streams} have proposed a domain-specific streaming language for Java as a library that was faster in some cases than the native Java streaming API. More generally though there seems to be little recent published work on streaming design patterns. We therefore provide a larger historical context on streaming or streaming-related languages by providing some older papers from the programming language community that are representative of ideas that are related. 

\textit{Stream processing in programming languages and dataflow programming.} A good overview of the earliest developments of stream processing from the 60s to the 90s has been written by Stephens~\cite{stephens1997survey}. It covers among others representative papers from the dataflow, functional, and logic approaches to stream programming, as well as reactive approaches and hardware design and verifications applications. From these different approaches we think the dataflow approach is closest to the pull-stream programming model. The first dataflow language is Lucid~\cite{wadge1985lucid}. In Lucid, variables are infinite streams and variable transformations, called \textit{filters}, are expressed as equations between variables. Filters therefore continuously compute new values based on the latest available values, similar to the pull-stream transformers. Lucid as a notation is more expressive than using pipelines of pull-stream modules: for example, it is easier in Lucid to express a complex dataflow network where let's say the output of a filter becomes in own input and is combined with multiple other input streams. 

Our own experience with stream processing with dedicated language support, which led to the insights we mentioned in Section~\ref{Section:Insights}, came from our experience with Oz~\cite{henz1993oz,van2004concepts}, a multi-paradigm language built around a core language of orthogonal language features which together provide support for all the major programming paradigms. In Oz, stream programming is done with explicit streams of single-assignment dataflow variables. This makes the additional concurrency possibilities of the pull-stream protocol much easier to use in practice since the stream data structure takes care of the synchronization issues, which is not the case in JavaScript.

Another more recent survey from Johnston et al. details later development in the 90s and early 2000s on dataflow approaches, languages, visual dataflow programming~\cite{johnston2004advances}. Recently, the dataflow programming paradigm was revived around distributed programming with languages such as Agapia~\cite{paduraru2014dataflow}.

\textit{Functional reactive programming.} From another perspective, functional reactive programming~\cite{wan2000functional} also integrated similar streaming notions as a library rather than as programming language primitives, with the Functional Reactive Animation library and its an associated denotational semantics, as a prime example~\cite{elliott1997functional,elliott1998functional}. One key difference compared to the dataflow approaches we mentioned previously is in the treatment of time as a continuous quantity and the associated problem of sampling to realize animations with a discrete number of frames.

\textit{Formal specifications.} Our own expertise does not lie in formal specification and our knowledge of the existing literature is limited. It is therefore likely that our event-based protocol language is similar to prior existing work. Accordingly, we do not make any claim of originality on it and recognize that better alternatives to specify pull-stream modules may exist. Nonetheless, our language definition has shown to be \textit{sufficient} to describe both the pull-stream protocol and specify reference modules and \textit{concise enough} to make the paper self-contained. We believe its main contribution will be to make it easier for experts in the respective fields to use this paper as a case study for their own work on formal methods and suggest better notation alternatives that could be used to specify pull-stream modules. 

That being said, our event-based protocol language uses predicate calculus~\cite{enderton2001mathematical} and models time implicitly by extending a history in a discrete step for each new event added. We were inspired to formalize pull-streams around events after reading existing work on formalizing distributed systems~\cite{Broy92thedesign,2014arXiv1409.7236K}. A state machine-based formalism could have been another valid alternative. Lamport's Temporal Logic of Action~\cite{lamport1994temporal} and its associated language TLA+~\cite{lamport1999specifying} are a foundation to reason about concurrent programs with a strong mathematical foundation which could have been another alternative. We do not have experience in using either but the introduction of TLA+~\cite{lamport1999specifying} claims that it is well suited to specify discrete asynchronous systems. This suggests they would be a good formal basis for our semantics. Another possible option would be temporal logic~\cite{pnueli1977temporal,clarke1986automatic}.

\textit{Automated testing.} Our testing strategy, is similar to the property-based testing approach of Claessens and Hughes~\cite{claessen2011quickcheck} in which a large number of test cases are automatically generated and some properties are checked at run-time to hold over all test cases. However, in contrast to their approach, we systematically test all cases for small finite streams rather than performing random testing. This is especially effective in our case as the behavior of pull-stream modules can often be defined inductively: testing for the base cases and a few inductive ones is usually enough to uncover most bugs. Another difference is that we do not test the module for functional correctness (i.e. that their output is correct given their input), we test them to ensure that whatever event they generate, their sequence correctly follow the pull-stream protocol. Our testing approach therefore focuses on \textit{interoperability} between modules.  

\textit{Others.} Recent work in the programming language community has focused on stream processing. Vaziri et al. have shown an approach based on extending spreadsheets with stream support and formalized the core of their language extension~\cite{vaziri2014stream}.

%% file: conclusion.tex
In this paper, we provided a formal treatment of the pull-stream design pattern that originated within the JavaScript developer community. We provided a new insight that the effectiveness of this design pattern comes from the \textit{declarative concurrent programming model} that it uses. We then presented an event-based protocol language that is independent of the original JavaScript implementation, which we then used to formally and concisely specify the pull-stream callback protocol and reference modules that generate all possible events.  This formalization ultimately led to new pull-stream module implementations in JavaScript that in turn helped automatically identify some corner-cases and possible inconsistencies in actual implementations of the most widely used modules. Our approach therefore helps to better understand the pull-stream protocol, ensure interoperability of community modules, and concisely and precisely specify new pull-stream abstractions in papers.

We envision many direct applications to our work: (1) test more community modules for conformity; (2) implement the event-based protocol  language in JavaScript and compare the execution of the specification to the implementation for consistency; (3) extend the run-time checking approach to test all possible inter-leavings of concurrent executions; (4) provide an ascii-based equivalent notation to document the behavior of community modules; (5) document other stream protocols using a similar approach. 

In addition, the pull-stream protocol itself is relatively simple and could serve as a great case study for ensuring that the latest verification techniques based on type checking are expressive enough to be applicable to protocol compliance. 


%% file: appendix.tex
\section{Normalization Rules}
\label{Section:Normalization-Rules}

Input for the proof of termination with Aprove:
\begin{verbatim}
(VAR x y e e1 e2 a)
(RULES
    before(e,and(x,y)) -> and(before(e,x),before(e,y))
    before(and(x,y),e) -> and(before(x,e),before(y,e))
    before(e,or(x,y))  -> or(before(e,x),before(e,y))
    before(or(x,y),e)  -> or(before(x,e),before(y,e))
    before(e,empty)    -> e
    before(empty,e)    -> e
    before(before(e1, empty), e2) -> before(e1, e2)
    before(e1, before(empty, e2)) -> before(e1, e2)
    and(a, empty) -> a
    and(empty, a) -> a
    or(a, empty) -> a
    or(empty, a) -> a  
)
\end{verbatim}

\section{Concurrent Variations of the Pull-Stream Protocol}
\label{Appendix:Concurrent-Variations}
\input{concurrent-variations}

%% file: concurrent-variations.tex
The pull-stream protocol could have used additional concurrency while keeping the same function signature for modules. We detail here the two additional cases that would have been possible but were ultimately ruled out because they incurred additional complexity in the implementation of modules to handle the concurrency while the benefits were not worth it. We present the two cases both for the \textit{normal sequence} and the \textit{early-termination sequence}.

\subsection{Normal Sequence}
\label{Section:Concurrent-Normal-Sequence}

There are two additional cases. The first case correspond to \textit{concurrent} asks with \textit{in-order} answers. The downstream module asks for multiple values, and the upstream module answers in the same order as the asks:
\[
\begin{aligned}
I: ask[\bar x_1], \quad I: ask[\bar x_2], \quad O: \bar x_1 := v_1, \quad O: \underline{terminated}(\bar x_2)
\end{aligned}
\]

The second case corresponds to \textit{concurrent} asks with \textit{out-of-order} answers. The downstream module asks for values concurrently and the upstream module answers in any order. This can be used to ensure minimum latency as answers are made as fast as they are available: 
\[
\begin{aligned}
I: ask[\bar x_1], \quad I: ask[\bar x_2], \quad O: \underline{terminated}(\bar x_2), \quad O: \bar x_1 := v_1
\end{aligned}
\]

In these two additional cases, the ordering of the stream values is still captured by the order in which the $asks$ were made and the downstream module can remember in which order the stream variables ($\bar x_1$ and $\bar x_2$) were created. While the additional implementation complexity is usually not worth it in JavaScript, these possibilities are actually quite natural to use in a dataflow language such as Oz because the stream of variables is explicitly represented in memory and all the synchronization happens implicitly as the values are bound to the variables.

\subsection{Early-Terminated Sequence}
\label{Section:Concurrent-Early-Termination-Sequence}

Similar to the normal sequence analysis, there are two additional cases to consider: the \textit{concurrent in-order} and the \textit{concurrent out-of-order} cases. 

The early-termination sequence can already issue a \textit{terminate} request concurrently with a single \textit{ask} request, so it can be considered as a limited of in-order concurrency. In the \textit{full concurrent in-order} case, more than one \textit{ask} are initiated concurrently before the \textit{terminate} request and therefore the \textit{terminated} answers are also received in-order. For example:
\[
\begin{aligned}
& I: ask[\bar x_1],~ I: ask[\bar x_2],~I: \underline{terminate}(\bar x_3),~O: \underline{terminated}(\bar x_1),~O: \underline{terminated}(\bar x_2),\\
& \quad O: \underline{terminated}(\bar x_3)\\
\end{aligned}
\]

The \textit{concurrent out-of-order} case is similar to the \textit{in-order} case except that answers may be received out-of-order which also applies to the termination answers. For example:
\[
\begin{aligned}
& I: ask[\bar x_1],~ I: ask[\bar x_2],~I: \underline{terminate}(\bar x_3),~O: \underline{terminated}(\bar x_3),~O: \underline{terminated}(\bar x_1),\\
& \quad O: \underline{terminated}(\bar x_2)\\
\end{aligned}
\]